\documentclass{emulateapj}

\usepackage{epstopdf}

\def\gs{\mathrel{\raise0.35ex\hbox{$\scriptstyle >$}\kern-0.6em
\lower0.40ex\hbox{{$\scriptstyle \sim$}}}}
\def\ls{\mathrel{\raise0.35ex\hbox{$\scriptstyle <$}\kern-0.6em
\lower0.40ex\hbox{{$\scriptstyle \sim$}}}}

\def\spose#1{\hbox to 0pt{#1\hss}}
\def\simlt{\mathrel{\spose{\lower 3pt\hbox{$\mathchar"218$}}
     \raise 2.0pt\hbox{$\mathchar"13C$}}}
\def\simgt{\mathrel{\spose{\lower 3pt\hbox{$\mathchar"218$}}
     \raise 2.0pt\hbox{$\mathchar"13E$}}}

\newcommand{\um}{\,$\mu$m\,}

\newcommand{\spitz}{{\sl Spitzer}}

\newcommand{\lir}[1]{10$^{#1}$\,$\rm{L}_{\odot}$}
\newcommand{\lsun}{\,$\rm{L}_{\odot}$}
\newcommand{\msun}{\,$\rm{M}_{\odot}$}

\newcommand{\z}{redshift\ }

\slugcomment{\it To be submitted to the Astrophysical Journal}

\shorttitle{Far-IR-to-radio properties of {\sl Spitzer}-selected high-$z$ ULIRGs }
\shortauthors{Sajina et al.}

\begin{document}

\title{{\sl Spitzer} Mid-Infrared Spectroscopy of  Infrared Luminous Galaxies at $z$\,$\sim$\,2 III:  \\
Far-IR to Radio Properties and Optical Spectral Diagnostics}

\author{Anna Sajina\altaffilmark{1}, Lin Yan\altaffilmark{2}, Dieter Lutz\altaffilmark{3}, Aaron Steffen\altaffilmark{2},\\
George Helou\altaffilmark{2}, Minh Huynh\altaffilmark{2}, David Frayer\altaffilmark{4}, Philip Choi\altaffilmark{5}, Linda Tacconi\altaffilmark{3}, Kalliopi Dasyra\altaffilmark{2}}

\altaffiltext{1}{Haverford College, Haverford, PA, 19041}
\altaffiltext{2}{\spitz\ Science Center, California Institute of Technology, Pasadena, CA 91125}
\altaffiltext{3}{Max Planck-Institut f\"ur extraterrestrische Physik, Garching, Germany}
\altaffiltext{4}{NASA {\sl Herschel} Science Center, California Institute of Technology, Pasadena, CA, 91125}
\altaffiltext{5}{Pomona College, Claremont, CA}

\begin{abstract}
We present the far-IR, millimeter, and radio photometry as well as optical and near-IR spectroscopy of a sample of 48 $z$\,$\sim$\,1\,--\,3 {\sl Spitzer}-selected ULIRGs with IRS mid-IR spectra. Our goals are to compute their bolometric emission, and to determine both the presence and relative strength of their AGN and starburst components.  We find that strong-PAH sources tend to have higher MIPS160\um\ and MAMBO\,1.2\,mm fluxes than weak-PAH sources. The depth of the 9.7\um\ silicate feature does not affect MAMBO detectability. We fit the far-IR SEDs of our sample and find an average $\langle L_{\rm{IR}}\rangle$\,$\sim$\,7\,$\times$\,$10^{12}$\lsun\ for our $z$\,$>$\,1.5 sources.  Our spectral decomposition suggests that strong-PAH sources typically have $\sim$\,20\,--\,30\% AGN fractions of $L_{\rm{IR}}$. The weak-PAH sources by contrast tend to have $\gs$\,70\% AGN fractions, with a few sources having comparable contributions of AGN and starbursts. The optical line diagnostics support the presence of AGN in the bulk of the weak-PAH sources.  With one exception, our sources are narrow-line sources, show no obvious correspondence between the available optical extinction and the silicate feature depth, and, in two cases, show some evidence for outflows. Radio AGN are present in both strong-PAH and weak-PAH sources. This is supported by our sample's far-IR-to-radio ratios ($q$) being consistently below the average value of 2.34 for local star-forming galaxies. We use survival analysis to include the lower-limits given by the radio-undetected sources, arriving at $\langle$\,$q$\,$\rangle$\,=\,2.07\,$\pm$0.01 for our $z$\,$>$\,1.5 sample.  In total, radio and, where available, optical line diagnostics support the presence of AGN in 57\% of the $z$\,$>$\,1.5 sources, independent of IR-based diagnostics. For higher-$z$ sources, the AGN luminosities alone are estimated to be $>$\,$10^{12}$\lsun, which, supported by the available [O{\sc iii}] luminosities, implies that the bulk of our sources host obscured quasars. 
\end{abstract}


\section{Introduction}

Ultraluminous Infrared  Galaxies (ULIRGs; defined as having $L_{\rm{IR}}$\,$>$\,\lir{12}, \citep{soifer87,sm96,genzel98,lonsdale06}) are believed to be an important contributor to the global star-formation rate density at $z$\,$\sim$\,2 \citep{lefloch05,caputi07}. This simple luminosity-cut definition, however, potentially includes a wide range of physical characteristics. It is particularly poorly understood at the high luminosity ($L_{\rm{IR}}$\,$\sim$\,$10^{12.5}$\,--\,$10^{13}$\lsun) end, due to the scarcity of such sources locally. In this regime and above, most sources are believed to be largely AGN-dominated \citep{lutz98,tran01,farrah02}. However, at $z$\,$\sim$\,2, sub-mm galaxies (SMGs) fall in the above luminosity range, but appear to be largely starburst-dominated ULIRGs \citep{alexander05,valiante07,pope07}. Although, less numerous, luminous quasars also peak at $z$\,$\sim$\,2 \citep{richards06}, and their bolometric luminosities fall in the above luminosity range. A popular evolutionary scenario suggests that major mergers lead to starburst-like cold ULIRGs followed by warm ULIRGs as the dust-obscured AGN become dominant. Eventually, outflows shed the dusty cocoon to reveal the naked quasar \citep{sanders88}.  Due to their mid-IR brightness {\sl Spitzer} Space Telescope \citep{werner04} selected $z$\,$\sim$\,2 ULIRGs, are biased to the `warm ULIRG' phase \citep{yan04,yan05,houck05} and so fill the gap between SMGs and QSOs allowing us to test the above scenario directly at the peak epoch for ULIRGs and quasars.  If at least a part of the quasars follow a similar evolutionary path we expect to find a population of obscured quasars which are accompanied by strong starburst activity. Recent studies suggest this might indeed be the case \citep{polletta07,ams07}. 

This paper concerns a sample of 52 {\sl Spitzer}-selected ULIRGs, with Infrared Spectrograph \citep[IRS;][]{houck04} spectra. By selection, our sample has somewhat less extreme 24\um-to-R colors compared with other samples aiming at understanding the {\sl Spitzer}-bright ($\gs$\,1\,mJy at 24\um), optically-faint population \citep[e.g.][]{houck05,weedman06}. Therefore, beside the strong continuum, strong silicate absorption spectra typical of such samples, our sample also includes sources with PAH features of comparable strength to those seen in SMGs  \citep{pope07}.  Our sample may therefore represent transition objects spanning the range between (un)obscured quasars and SMGs. This is PaperIII of a series of papers on the infrared properties this sample. In \citet[][hereafter Paper I]{yan07}, we presented the spectra, sample selection as well as the IRS-derived redshifts. In \citet[][hereafter PaperII]{me07}, we presented the spectral decomposition of the IRS spectra into their PAH, continuum, and obscuration components. 

An {\sl HST} NICMOS morphology study of the 36 $z$\,$>$\,1.5 sources in our sample is presented in \citet{dasyra08}.   That study shows that, typically, the most compact sources are those with AGN-like mid-IR spectra (i.e. weak PAH) and vice versa.  The more intriguing result was the finding that at least some of the interactions leading to ultraluminous activity at $z$\,$\sim$\,2 appear to differ from local ones (which essentially require equal mass mergers).  This result cautions that some of our sources may not have true local analogues. In \citet{rl_letter}, we discussed the fact that 40\% of our $z$\,$>$\,1.6 sources with optical depths of $\tau_{9.7}$\,$>$\,1 are radio loud, including some sources with powerful radio jets. Since deep silicate feature sources are particularly prone to power source ambiguities (due to the heavy obscuration), this result was crucial in confirming that a significant fraction of these sources do indeed host powerful AGN.   Lastly, the {\sl Chandra} X-ray detectability of our sources is addressed in Bauer et al. (2008, in prep.).  There are however key questions about the nature of these sources which remain unanswered after the above studies, and are the subject of this paper. Given the uncertainties inherent in the mid-IR diagnostics (see PaperII), can we confirm the presence of obscured AGN in our sources, and if present, are they of quasar-like strength? What is the bolometric power output of our sources, and what fraction of that is due to AGN vs. star-formation activity?

This paper is organized as follows.  \S\,2 presents the available multiwavelength data and data reduction including Keck and Gemini optical and near-IR spectroscopy, Multiband Imaging Photometer for Spitzer \citep[MIPS;][]{rieke04} 70\um\, and 160\um, and MAMBO\,1.2\,mm pointed photometry, as well as the archival VLA 1.4\,GHz, and GMRT 610\,MHz data.  Our analysis of the above data is presented in \S\,3, where we look for the signature of AGN activity via their radio properties (luminosity and spectral index) as well as traditional optical line diagnostics (broad lines, high ionization lines, and AGN-like line ratios). We use spectral decomposition to determine the role of AGN and starburst activity in powering the bolometric output of our sources.  In \S\,4, we use all of the above diagnostics to address the question how do our sources relate to other high-$z$ luminous populations such as SMGs and in particular whether or not they form part of the population of obscured quasars as we speculated in PaperI. Finally, we summarize the key results of this work in \S\,5. Throughout this paper, we adopt a $\Lambda$CDM model with $\Omega_{\rm{M}}$\,=\,0.27, $\Omega_{\Lambda}$\,=\,0.73, and $H_0$\,=\,71\,$\rm{km}\rm{s}^{-1}\rm{Mpc}^{-1}$ \citep{spergel03}.

\section{Data \label{sec_data}}
\subsection{IRS spectral sample}

PaperI gives the detailed description of the sample selection, presents the observed spectra and \z measurements.  Here we summarize the key points. Our sample is selected from the 4 sq.deg.\ {\sl Spitzer} Extragalactic First Look Survey (XFLS\footnote{\url{http://ssc.spitzer.caltech.edu/fls/}}). The full sample consists of 52 sources  with 24\um\ flux density brighter than 0.9mJy, and 
red 24-to-8\um\ and 24-to-R colors\footnote{(1) $R(24,8) \equiv \log_{10}(\nu 
f_{\nu}(24\mu m)/\nu f_{\nu}(8\mu m) \simgt 0.5$; 
(2) $R(24,0.7) \equiv \log_{10}(\nu f_{\nu}(24\mu m)/\nu f_{\nu}(0.7\mu m) 
\simgt 1.0$.}. The IRS low resolution data revealed that 47 of the targets
have measurable redshifts from the features detected in their
mid-IR spectra, with the majority (35/47\,=\,74\%) at 1.5\,$<$\,$z$\,$<$3\,.2, 
and a smaller fraction (12/47\,=\,28\%) at 0.65\,$<$\,$z$\,$<$\,1.5. An additional source (MIPS279) has a Keck redshift bringing our sample to 48.  Most of our spectra cover rest-frame $\sim$\,5\,--\,15\um\ with a minimum signal-to-noise per pixel of 4. Because the primary selection is the 24\um\ flux, we name our sources based on their MIPS24\um\ catalogue number (see Table~\ref{table_long} for a source list and redshifts). 

In PaperII, we presented the decomposition of the rest-frame $\sim$\,5\,--\,15\um\ mid-IR spectra into PAH, continuum, and obscuration components.  Based on the 7.7\um\ PAH feature equivalent widths, we define `strong-PAH' sources as those having EW7.7\,$>$\,0.8\um, while `weak-PAH' sources have EW7.7\,$<$\,0.8\um.  We adopt these definitions throughout this paper as well. In PaperII, we found that $\sim$\,75\% of the sample are weak-PAH sources consistent with AGN-dominance in the mid-IR. However, although of comparable luminosities, these sources are more obscured than typical  quasars as evidenced by their steeper slopes, and typically significant 9.7\um\ silicate absorption. Although most of our strong-PAH (starburst-like) sources were  found among the lower-$z$ sub-sample, we found an intriguing population of these star-forming sources at $z$\,$\sim$\,2 with luminosities  an order of magnitude larger than local ULIRG analogues. 

\subsection{Optical spectroscopic observations \label{sec_opt_specs}}

We obtained optical spectra for 12 of our sources using the LRIS and DEIMOS instruments on the Keck telescopes \citep{oke, faber}.  The initial results from the optical and near-IR spectroscopy (see below) were presented in PaperI \citep{yan07}.   For the red side of the LRIS spectrograph, he D680 dichroic was used, splitting the red and blue arms at 6800\AA. The red arm was equipped with the 400 line mm$^{-1}$ grating blazed at 8500\,--\,10000\AA. At the blue side, we used the 300 line mm$^{-1}$ grating blazed at 5000\AA. Roughly, the full spectral range is 3200\,--\,9600\AA. The slit width is $1.2^{''}$ and we dithered the targets along the slits between exposures with offsets of 2\,--\,4$^{''}$.  The integration time for each exposure was 1200\,seconds, with a total time per mask of 1\,--\,1.5\,hours. The DEIMOS observations were made using a 1200 line mm$^{-1}$ grating with central wavelength settings of 7400\AA\ and 7699\AA\ together with the GG495 blocking filter, resulting in a 0.33\AA\ pix$^{-1}$ mean spectral dispersion and a 1.45\AA\ instrumental resolution. The wavelength coverage for DEIMOS spectra is 6300\,--\,9300\AA. The total integration time for DEIMOS observations is 3,600\,seconds per mask. 

For the LRIS data reduction, we used a suite of IRAF scripts written by D. Stern (private communications). The basic procedure is described in detail in \citet{yan04aj}. For the DEIMOS data, we used the DEEP2 {\sc spec2d} data reduction pipeline, which performs cosmic ray removal, flat-fielding, co-addition, sky subtraction, wavelength calibration and both 2-d and 1-d spectral extraction. 

\subsection{Near-IR spectroscopic observations  \label{sec_specs}}
We obtained near-IR $H$ and $K$ spectra using the NIRSPEC instrument on Keck \citep{mclean}. Additional three nights with the NIRI instrument \citep{hodapp} on the GEMINI-north telescope were used to obtain near-IR $H$ and $K$-band spectra of our sample targets at $z\sim 1.8-2.4$.
Near-IR spectra with sufficient signal-to-noise ratios were obtained for 11 sources. The NIRSPEC observations were done in moderate resolution of $\sim1500$, with a slit of $42^{''}$ in length and $0.76^{''}$ (4pixels) in width.  At this resolution, the NIRSPEC instrument FWHM is $\sim$\,10.0\,\AA ($\sim$\,200\,km/s) at $H$ and $\sim$\,14\,\AA ($\sim$\,190\,km/s), as determined from isolated sky lines \citep[see also][]{brand07}. The NIRI observations were taken with the f/6 camera, with a field of view of 120$^{''}$\,$\times$\,120$^{''}$.  We selected the slit width of $0.75^{''}$ (6.42\,pixels) and achieved roughly $520$ spectral resolution in H (1.43-1.96\um) and K-band (1.90\,--\,2.49\um).  The NIRI instrument FWHM is $\sim$\,20\,\AA ($\sim$\,400\,km/s) at $H$ and $\sim$\,28\,\AA ($\sim$\,380\,km/s) in $K$ \citep{brand07}. The sources for the near-IR spectroscopy were chosen to have appropriate redshifts so that their expected H$\alpha$ and [O{\sc iii}] emission lines are within the observed $H$ or $K$-band. The total on-target integration time per source ranges from 45minutes to 1.5\,hours, depending on the strength of the emission lines and the observing conditions.  

\subsection{Near-IR data reduction and flux calibration \label{sec_calib}}

The NIRSPEC data reduction was done using IDL software written by
George Becker (private communication). This software uses the methods
of optimal sky subtraction \citep{kelson03} for reducing longslit spectra
in the near-infrared, where the strong OH emission lines can hide the
faint spectral features of our source sample. This method models and
removes sky emission without first rectifying or rebinning the data,
which eliminates noise and edge effects that can occur during
interpolation. The spectra are wavelength calibrated using the sky
emission lines. Standard F and G stars were observed in order to
perform flux calibrations and to remove atmospheric absorption
features. The absolute standard fluxes were obtained from the IRTF
Spectral Library\footnote{http://irtfweb.ifa.hawaii.edu/~spex/spexlibrary/IRTFlibrary.html}
using the spectral type closest to that of our standards. The maximum air mass difference between our source observations and the standard star observations is $\sim$\,0.2. Given the average values of sky extinction on Mauna Kea, which are 0.05\,mag/airmass, and 0.07\,mag/airmass in the $H$ and $K$ band respectively \citep{maunakea_ext}, this airmass difference does not significantly affect our flux calibration. 

The aforementioned IDL code for longslit reductions does not currently
accept NIRI data, so we performed the spectroscopic reductions on the
NIRI data using the Gemini IRAF package GNIRS and the
instrument-specific NIRI package, following the data reduction
procedures outlined on the Gemini Observatory
webpages\footnote{http://www.gemini.edu/sciops/instruments/niri/data-format-and-reduction}.
The flux calibration followed a similar procedure as for NIRSPEC. The spectra were extracted using a Gaussian profile with FWHM of 4pixels for NIRSPEC and 8pixels for NIRI, appropriate for spatially unresolved sources. 

Lastly,  the absolute flux correction is uncertain due to the  uncertain positioning of the standard sources on the slits and possible light loss \citep[see also][]{brand07}. In order to estimate the degree of uncertainty in the absolute fluxes, we compare with expectations from the NICMOS broadband photometry \citep{dasyra08}. This is complicated for most of our sources due to the non-detection of the continuum. The only exception is MIPS8196 (observed with NIRI), which is brighter than the rest of our  sources ($m_{H,\rm{Vega}}$\,=\,18.68).  We find that, convolving its spectrum with the NICMOS F160W transmission curve results in a flux 1.6 times the observed. This sources also has a $K$\,-\,band magnitude from the Wide Field IR Camera (WIRC), of $m_{K,\rm{Vega}}$\,=\,17.63 (P.Choi, private communication). Our $K$-band spectrum results in a WIRC flux a factor of 1.9 higher than the observed.  This shows that the absolute flux calibration of our sources is likely overestimated by roughly 50\,--\,100\%. However, as we cannot confidently determine these overestimation factors for any other sources (both NIRSPEC and NIRI), we choose not to apply this correction factors, but merely state that the line luminosities we derive from these spectra (\S\,\ref{sec_lineratios} and Table~\ref{table_keck}) are uncertain up to a factor of 2 (0.3\,dex). 

\subsection{MIPS 70\um\ \label{sec_mips70}}

The XFLS was observed in scan map mode with the MIPS70\um\ array \citep{frayer06}. A smaller, 0.25 sq.deg., verification field was imaged with 4 times the integration time as the main XFLS field.  The noise is non-uniform across the field, however the average point source (1\,$\sigma$) rms at 70\um\  is $\sim$\,2.8\,mJy in the main field, and $\sim$\,1.6\,mJy in the verification field.  Three of our sources fall inside the verification field (MIPS15977, MIPS8268, MIPS8245). 
For the sources without clear 70\um\ detection in the scan map data, additional pointed photometry data was obtained as part of {\sl Spitzer} GO2 program \#20239 (PI: P.Choi). Most of our targets have roughly 700 seconds integration time per pixel.  The scan map data and pointed photometry data were combined into a local median-subtracted image with 4\arcsec\ pixel scale.  

The fluxes and their uncertainties are estimated from PRF photometry on the median-subtracted image \citep[using APEX;][]{makovoz05}.  For color corrections, we use the values for flat spectrum ($F_{\nu}$\,$\propto$\,$\nu^{-1}$), which is also roughly consistent with the color corrections for dust temperatures in the range 30\,--\,70\,K.  This implies a color correction of 1.09 for MIPS70\um, and 1.04 for MIPS160\um (see below). An additional 5\% error in the flux is attributed to the uncertain color correction (after excluding obviously inconsistent spectral types). This uncertainty is negligible given the generally low signal-to-noise of our data.  The typical rms flux error is $\sim$\,1.7\,mJy.  

We detected 29 of the 48 sources listed in Table~\ref{table_long}, based on a 10\arcsec\ matching radius (the positional offsets and possibility of random matches are addressed in \S\,\ref{sec_random}).   The aperture-photometry derived 70\um\ fluxes for our sample were presented in PaperII. We found that, within the expected uncertainty the PRF-derived fluxes presented here and these aperture-photometry derived fluxes agree well, however, that the PRF method is somewhat more conservative in that 4 sources with low signal-to-noise detection in PaperII (SNR\,$\sim$\,3\--\,3.5) were not detected with APEX. Here we adopt the PRF fluxes as being more conservative. 

\subsection{MIPS 160\um\ \label{sec_mips160}}

The XFLS field and verification field were also imaged with the MIPS160\um\ array \citep{frayer06}. The noise is quite non-uniform but typically $\sim$\,20\,mJy in the main field, and $\sim$\,10\,mJy in the verification field. Two of our sources are in the MIPS160\um\ catalogue: MIPS22404 and MIPS15977 (which is in the verification field).  Our sample, except MIPS22404, was also observed in 160\um\ pointed small-field photometry ({\sl Spitzer} GO2 program \#20239 as above). The total integration time per pixel is estimated to be $\sim$\,2500\,sec.  

The fluxes and their uncertainties are estimated from PRF photometry on the median-subtracted images as in the case of the MIPS70\um\ data.  We used a 20\arcsec\ matching radius.  The positional offsets and possibility of random matches are addressed in \S\,\ref{sec_random}. We adopt the latest value of the MIPS160\um\ calibration, 41.7\,MJy/ster \citep{mips160_calib}. The MIPS\,160\um\ calibration is uncertain to $\sim$\,15\%. As mentioned in \S\,\ref{sec_mips70}, we adopt a color correction of 1.04 for MIPS160\um.  An additional 10\% correction is needed, since for small field photometry filtering  results in an underestimated flux densities (see MIPS Handbook).  We confirm that this is present in our data by comparing the pixel values at the positions of a few of the brightest sources (including MIPS15977, which is detected in both the field map and the photometry data).  Therefore,  all APEX-derived fluxes are multiplied by 1.14, which includes color correction, as well as the above offset.  A total of 9 sources have $>$\,3\,$\sigma$ detections (1 more has 2.6\,$\sigma$). Including MIPS22404, this adds up to 11 sources (23\% of the sample). For the rest, we use the 3\,$\sigma$ upper limits in all analysis.  The typical 1\,$\sigma$ uncertainty of our MIPS\,160\um\ data is $\sim$\,9\,mJy. 

\begin{figure}[!ht]
\begin{center}
\plotone{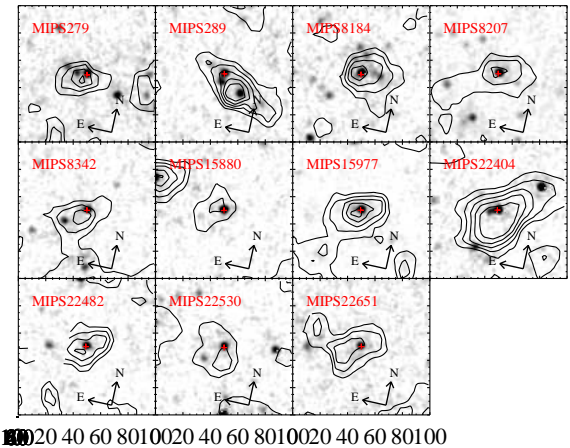}
\end{center}
\caption{The 11 160\um-detected sources (our IRS sources are marked with small red crosses). Here we combine the XFLS map with our targeted photometry data. The 24\um\ image is overlaid with the MIPS160\um\ 1\,--\,5\,$\sigma$ contours.   In some cases, multiple 24\um\ sources contribute to the 160\um\ emission (see \S\,\ref{sec_multi} for details).  The boxes are 120\arcsec\,$\times$\,120\arcsec, with the N-E orientation as shown. \label{mosaic_full}}
\end{figure}

As an illustration of our 160\um\ data, Figure~\ref{mosaic_full} shows the MIPS 24\um\ images, centered on our target and overlaid with the 160\um\ contours. We show the 11 sources with highest signal-to-noise ratio at 160\um, where we combined the XFLS map \citep{frayer06} with our photometry data.  An important point we see in Figure~\ref{mosaic_full} is that in some cases, the shape of the 160\um\ contours (e.g. MIPS289, MIPS8184, MIPS22404)  suggests that more than one source contributes significantly.  We address this issue in \S\,\ref{sec_multi}.

\subsection{MAMBO 1.2\,mm}
Observations were carried out at the IRAM 30\,m telescope using the 117 element version of the Max Planck Millimeter Bolometer (MAMBO) array (Kreysa et al. 1998) operating at a wavelength of 1.2\,mm, during several pool observing periods between October 2004 and March 2007. On-off observations were typically obtained in blocks of 6 scans of 20 subscans each, and repeated in later observing nights until reaching a total rms near 0.66\,mJy or a 5\,$\sigma$ detection. The median 1.2\,mm 1\,$\sigma$ rms noise level reached for the sample is 0.59\,mJy (range 0.35\,--\,0.82 mJy).  The data were taken during pool periods with good atmospheric transmission $\tau_{1.2\,mm}\lesssim 0.3$ and low or medium skynoise, for a typical total observing time of 2h per source spread over several periods. Data were reduced with standard procedures in the MOPSIC package developed by R. Zylka. The flux calibration is based on multiple observations of the IRAM flux calibrators during the pool sessions and
monitoring of the 1.2\,mm atmospheric transmission through regular skydips, and is believed to be accurate to $\sim$15\%. Correlated sky noise was subtracted from the active channel using average signals from surrounding channels. All data taken during a several weeks period affected by
technical problems after a change of the telescope control system were discarded, as well as individual scans likely subject to telescope or instrument problems according to the telescope log and inspection of the data themselves.

MAMBO 1.2\,mm observations have been obtained for 44/48\,=\,92\% of our IRS sample with redshifts.  Of these, 7 are detected at $>$\,3\,$\sigma$ (14 at $>$\,2\,$\sigma$).  All sources without individual 1.2\,mm detections, have a weighted mean signal of $\langle S_{\rm{1.2mm}}\rangle$\,=\,0.47\,$\pm$\,0.09\,mJy (the uncertainty is error-in-mean, i.e. $rms/\sqrt{N}$), and are therefore detected as a sample at $>$\,5\,$\sigma$ significance.

Part of these observations were previously presented in \citet{lutz05}. \citet{lutz05} obtained MAMBO photometry for a sample of bright XFLS 24\um sources with $R$(24,0.7) similar to our IRS sample, but without the additional $R$(24,8) constraint. Including only targets from our IRS sample, Table\,\ref{table_long} lists all available 1.2\,mm fluxes and their statistical uncertainties.  

\subsection{VLA 20\,cm}

Radio 1.4\,GHz (20\,cm) fluxes are available from the Very Large Array (VLA) map of the XFLS field \citep{condon03}.  This was done in B-array with a FWHM of 4.8\arcsec\ at 1.4\,GHz. The typical rms of the map is $\sim$\,23\,$\mu$Jy. We matched our sources to the 3.9$\sigma$ ($S_{\rm{1.4GHz}}$\,$>$\,90\,$\mu$Jy) catalog (J.Condon, private communication).  We assumed a matching radius of 2.5\arcsec\ (the size of the MIPS24\um\ beam). We find counterparts for 30 of our sources, where the positional offset with respect to the 24\um\ positions was in the range 0.2\arcsec\,--\,1.6\arcsec\ with an average value of 0.9\arcsec. One additional source, MIPS8342 is included despite having a positional offset of 3.6\arcsec\ as its radio image looks like this might be an extended source associated with the 24\um\ source. The smaller XFLS verification field was also imaged to an rms of $\sim$\,8\,$\mu$Jy using the Westerbork Synthesis Radio Telescope (WSRT) \citep{morganti04}. Only 3 of our sources (MIPS8245, MIPS8268, and MIPS15977) fall in this field. For these, we use the deeper observations (MIPS8268 is undetected in the shallow map).  This brings our radio detections to 67\% of the sample. For the rest, we use 3\,$\sigma$ upper limits. The 1.4\,GHz radio fluxes are listed in Table~\ref{table_long}. 

\subsection{GMRT 610\,MHz \label{sec_gmrt}}

In addition to the 1.4\,GHz fluxes, 48\% of our sample is detected in the Giant Meterwave Radio Telescope (GMRT) 610\,MHz map of the XFLS field \citep{garn07}.  The typical 3\,$\sigma$ limit of the catalogue is $\sim$\,0.3\,mJy. We describe the radio spectrum as a power law of the form $F_{\nu}$\,$\propto$\,$\nu^{\alpha}$, where $\alpha$ is the spectral index.  At the redshifts of our sample, the GMRT data (combined with the VLA data) allows us to constrain the rest-frame $\sim$\,1.4\,--\,5.0\,GHz radio spectral indices. The available 610\,MHz fluxes as well as the radio spectral indices are listed in Table~\ref{table_long}. Lastly, we would like to note that all fluxes (70\um\,--\,610\,MHz) shown in Table~\ref{table_long} represent total fluxes, therefore no further aperture matching is required.

\subsection{Effect of multiple sources \label{sec_multi}}

\begin{figure}[!h]
\begin{center}
\plotone{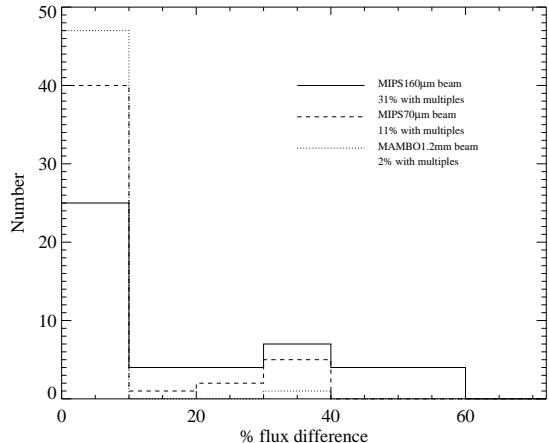}
\end{center}
\caption{The relevance of multiple 24\um\ sources within the MIPS70\um, MIPS160\um, and MAMBO1.2\,mm beams. The x-axis is calculated as $100(1-\frac{S_{24,max}}{\sum{S_{24}}})$. Our sources usually correspond to $S_{24,max}$. If the percentage flux difference is within $\sim$\,30\%, it is likely included in the far-IR flux error, and is negligible.  Differences $\gs$\,30\% are likely to have two or more IR sources significantly contributing to the measured flux.  \label{mips160_conf}}
\end{figure}

Our IRS spectra cover a spatial scale of $\sim$\,5\arcsec, comparable to the MIPS24\um\ beam.  By comparison the MAMBO1.2mm, MIPS70\um\ and 160\um\ beams are $\sim$\,11\arcsec, 20\arcsec, and 40\arcsec\ respectively.  The large beams, especially of MIPS160\um, suggest the possibility of multiple sources contributing to the measured flux (see also Figure~\ref{mosaic_full}). In Figure~\ref{mips160_conf}, we quantify this by measuring the percentage of the total 24\um\  flux within a given beam that is contributed by the brightest 24\um\ source.  This analysis has been applied previously in the case of  {\sl ISO} 170\um\ sources \citep{me06}, as well as SMGs \citep{pope06}.   

Given the low signal-to-noise ratio of our sources, we assume multiple sources are significant only where the percent flux difference is  $>$\,30\,\%. Within the MIPS160\um\ beam, 15 sources meet this criterion (31\% of the sample). Within the MIPS70\um\ beam, 5 sources have multiple 24\um\ sources (11\% of the sample). Finally, only  1 of our source has multiples within the MAMBO beam (2\% of our sample). For sources without individual detections, attempting to deblend multiple sources is futile. For the sources with individual 160\um\ and/or 70\um\ detections and multiple 24\um\ contributors, we discuss our deblending strategy on a source-by-source basis in \S\,\ref{sec_deblend}.  
  
\subsubsection{De-blending MIPS70\um\ and 160\um\ sources \label{sec_deblend}}
\noindent{\bf MIPS110} -- Here there is an additional 24\um\ source (0.8\,mJy) close to our source (1.8\,mJy). However, these sources are within 2\arcsec\ of each other (this is the only multiple source within a MAMBO beam).  This close, we cannot disentangle the contribution of the two 24\um\ sources and hence treat them as one. There is an offset between the 70\um\ APEX source and our nominal source position (7.2\arcsec); however, the 70\um\ peak does not correspond to any other 24\um\ source, and hence is still most likely due to our source (see e.g. \S\,\ref{sec_random}).   \\
{\bf MIPS279} -- This is a chain of three 24\um\ sources starting with our target. They are separated by $\sim$\,10\arcsec\ between the first and second and second and third sources. The 70\um\ source is centered on our target (positional offset 1.0\arcsec), and hence we assign the full 70\um\ flux to our source. Of the three, only our source has a 20\,cm radio detection. The radio flux is 6\,$\sigma$ the local map rms suggesting that the undetected radio emission of the other sources is significantly lower than our target's. Therefore most likely the MIPS160\um\ far-IR emission  is dominated by our target as well.  \\
{\bf MIPS289} -- Our source appears blended with a nearby bright 54\,$\pm$\,10\,mJy source ($\sim$\,20\arcsec\ away).   We attempt to deblend the two by combining two 2D Gaussians, and find that our source has about half the flux of its neighbor.  We therefore assign it a flux of $\sim$\,27\,$\pm$\,10\,mJy.   \\
{\bf MIPS8184} -- This source is part of a group of 6 MIPS24\um\ sources ranging from $S_{24}$\,=\,0.4\,--\,1.2\,mJy within the 160\um\ beam.  Our source position however is the strongest 70\um\ source. The ratio between the peak 70\um\ flux at the position of our source and the next brightest peak is 1.5. Therefore we assume that our source dominates the 160\um\ emission as well. \\
{\bf MIPS8196} -- Here there are three 24\um\ sources, one 5.2\arcsec\ from our target and the other a further $\sim$\,10\arcsec\ from the second source. The 70\um\ emission is centered in between and seems to be largely contributed to by the later two sources.  Therefore we treat this source as a 70\um\ non-detection. \\
{\bf MIPS8342} -- This system has a nearby ($\sim$\,20\arcsec) neighbor of roughly equal 24\um\ flux. The centroid of the 160\um\ emission lies roughly in-between the two 24\um\ sources (see Figure~\ref{mosaic_full}).  The ratio of the 70\um\ flux ratio between the position of our source and the secondary source is 1.2, which is also consistent with the two sources having comparable far-IR emission.  It seems reasonable to assume that our source contributes half of the total 160\um\ flux (29\,$\pm$\,11\,mJy). We therefore take its deblended 160\um\ flux to be half of that, or 14\,$\sim$\,11\,mJy.   \\
{\bf MIPS22404} At 160\um\, this is a large blob encompassing at least four 24\um\ sources (see Figure~\ref{mosaic_full}). Two of the sources are far enough away that they are excluded from the PRF-derived 160\um\ flux. However, the nearest source to our target is 7.2\arcsec\ away, with the 70\um\ emission centered roughly in between our source and its neighbour.  Moreover, both are radio sources. It is likely that both sources contribute to the far-IR emission, although their relative contribution is uncertain. For simplicity, we split both the 70\um\ and 160\um\ fluxes between our target and its neighbor. 

Due to the uncertain nature of the above deblending procedure, we prefer to quote the original APEX fluxes in Table~\ref{table_long}, but use the deblended values in all subsequent analysis. 

\subsection{Random matches \label{sec_random}}
In \S\,\ref{sec_mips70} and \S\,\ref{sec_mips160}, we discussed the flux determination for the MIPS70\um\ and MIPS160\um\ observations respectively. In both cases, we ran APEX first to find all sources in the images and then matched those against our 24\um\ source positions. Table~\ref{table_long} shows the fluxes of the detected sources as well as the positional offsets between the APEX-derived 70 and 160\um\ sources and the reference 24\um\ positions. In this section, we address the chances of random associations in this procedure for cases where there is a single 24\um\ source.  The cases, where multiple 24\um\ sources may contribute to the observed far-IR emission are discussed in \S\,\ref{sec_deblend}.  
 
We apply the $P$-statistic \citep{downes86}, which estimates the random chance of one or more sources being found within a radius $\theta$ of a given source. It is defined as $P$\,=\,1\,-\,$exp(-\pi n \theta^2)$, where $n$ is the source density above a given flux level, and $\theta$ is the search radius. Neglecting the cases where multiple 24\um\ sources can contribute to the 70\um\ flux we find that the positional offsets vary between 0.1\arcsec\ and 6.9\arcsec, with an average value of 2.4\arcsec. This is comparable to the nominal MIPS70\um\ positional uncertainty  which is 2.6\arcsec\ \citep[1\,$\sigma$;][]{frayer06}. We compute the P-statistic for the source, which has the highest chance to be a random match.  This is MIPS22600 where a 6.4\,mJy 70\um\ source is found at 6.9\arcsec\ from the 24\um\ source. We use the 70\um\ counts from \citet{frayer06_counts70} which give 2.6\,$\times$\,$10^6$gals/ster $>$\,6.68\,mJy (with a counts bin ranging from 6.0 to 7.5\,mJy). Given $\theta$\,=\,6.9\arcsec\ for MIPS22600, we find $P$\,=\,0.009.  Given that the faintest detected source (MIPS8242 with 4.8\,mJy) is already 75\% the flux of MIPS22600, but with smaller positional offset, the rest of the sample would have even lower probabilities for being random associations.  The low $P$ values mean that our 70\um\ detections are not random. 

For MIPS160\um\ we have 11 detections (see \S\,\ref{sec_mips160}); however, 5 of these are blended sources (see \S\,\ref{sec_deblend}) making this statistic difficult to apply. For the remaining 6 sources, we find that the positional offset between the 160\um\ and 24\um\ source ranges from 1.5\arcsec\ (MIPS22482) to 9.8\arcsec\ (MIPS22651), with an average offset of 6.3\arcsec. This is comparable to the typical positional accuracy of 5.2\arcsec\ at 160\um \citep[1\,$\sigma$;][]{frayer06}.  MIPS15880 represents the `least likely' detection having among the greatest separation between the 24\um\ and 160\um\ sources (9.3\arcsec) as well as the lowest flux (23.8\,mJy). For 160\um, there are no published source counts that go down to our lowest detected fluxes ($\sim$\,30\,mJy). Based on the model counts from \citet{dole04_counts}, the range of acceptable models gives  $\sim$\,1\,--\,3\,$\times$\,$10^{6}$gals/ster having $\gs$\,30\,mJy.  For MIPS15880, a 23.8\,mJy source within 9.3\arcsec, we find that the above range gives $P$\,=\,0.006\,--\,0.019.   The rest of the non-blended 160\um\ detections, would have $P$\,-values no worse than the above.  Therefore, we conclude that likely none of the 160\um\ sources are random associations of our 24\um\ sources. 

\subsection{Stacking 70\um\ undetected sources \label{sec_stack70}}
Of the weak-PAH, $z$\,$>$\,1.5 sources nearly half (12/27) are undetected at 70\um\ (see Table~\ref{table_det}). The average of the aperture photometry measurements for these sources (see Paper\,II) is $\sim$\,2\,mJy which is about four times lower than the average of the 70\um-detected weak-PAH sources at $z$\,$>$\,1.5 ($\sim$\,10\,mJy).  For another measurement of the average flux of these sources we also stack the 70\um\ image at the positions of these sources. Our stacking largely follows the technique of \citet{huynh07b}. Here we summarize the key points. Before stacking, we subtract all bright ($\ge$\,5\,$\sigma$) sources from the median-subtracted image. To estimate the reliability of our stacking result, we also perform 500 stacks of 24 random position. We find a stacked flux of 1.95\,mJy. However, the standard deviation of the offset stacks is 0.68\,mJy making this only a 2.8\,$\sigma$ detection. 
 
\subsection{Stacking 160\um\ undetected sources \label{sec_stack160}}
Table~\ref{table_det} shows that only 3 of the 27 $z$\,$>$\,1.5 sources are individually detected at 160\um. These are all also detected at 70\um.  There are a total of 24 $z$\,$>$\,1.5, weak-PAH sources without individual 160\um\ detections. However, we note that our 160\um-detected sources have fluxes already roughly consistent with the model confusion limit of 160\um\ \citep{dole04_confusion}. This suggests that stacking is unlikely to improve this significantly. Nevertheless, using the same stacking technique as describe in \S\,\ref{sec_stack70}, we arrive at an average 160\um\ flux of 9.2\,mJy. The standard deviation of the offset stacks is 5.9\,mJy. Therefore these sources are undetected in the stacked image. We adopt the 2\,$\sigma$ upper limit for these sources (i.e. 12\,mJy).  
It is likely that the 70\um\,-undetected among these sources have lower 160\um-fluxes than the 70\um-detected ones; however, we find that we cannot distinguish between these sub-samples in our stacking for the reasons described above. 

\section{Results \label{sec_results}}
\subsection{Overview of our AGN/starburst diagnostics}

There are two distinct questions we wish to address here: 1) whether or not our sources contain AGN and/or starbursts, and 2) what is the relative contribution of the two processes to their bolometric power output. 
The presence of AGN in the bulk of our sample was originally asserted on the basis of their typically low PAH equivalent widths, and high 6\um\ hot dust continuum (see PaperII).  Conversely, the strong PAH emission in a quarter of the sample, and the presence of PAH in the stacked spectra of the weak-PAH sources suggest that star-formation is likely present in the majority of our sources (PaperII).  The presence of AGN in many of the weak-PAH sources was independently shown on the basis of their AGN-like radio luminosities and even, in two cases, radio jets \citep{rl_letter}.  

In this section we further test for the presence of starbursts by the detectability of cold ($\ls$\,50\,K) dust emission typically associated with this process.  We further expand the discussion on the presence of radio AGN on the basis of radio luminosities, spectral indices, and the ir-radio correlation. For a sub-set of the sample with available optical/near-IR spectra, we present optical line diagnostics which could reveal the presence of AGN. 

Finally, we use SED decomposition to estimate the relative contribution of AGN and starbursts to $L_{\rm{IR}}$. To do so, we make the simplifying assumption that the cold dust continuum is entirely due to star-formation while the warm/hot dust continuum is entirely due to AGN. The derived starburst component luminosities are the {\it sum} of the cold dust and PAH luminosities. We compare our results from the above analysis with the SEDs of well known sources and models in order to assess the level of uncertainty in the relative AGN/starburst fractions of $L_{\rm{IR}}$. In particular, we address the questions: how much starburst activity is there in the weak-PAH sources and how much AGN activity is there is the strong-PAH sources?

\subsection{Far-IR continuum: starburst or obscuration? \label{sec_origin}}
Typically, the far-IR emission is assumed to arise from star-formation even in quasars \citep[e.g.][]{beelen06,marshall07,polletta07}.  However, for highly obscured AGN as is the case with our sources (see PaperII), there exists the possibility of dust reprocessing the typically hot/warm SED of an AGN into a much cooler starburst-like SED \citep[e.g.][]{levenson07}.  Therefore, before we proceed, we look for trends between the far-IR continuum  emission and the PAH equivalent widths (indicative of starburst activity) and 9.7\um\ silicate feature optical depth ($\tau_{9.7}$) indicative of obscuration. 

\begin{figure}[!ht]
\begin{center}
\plotone{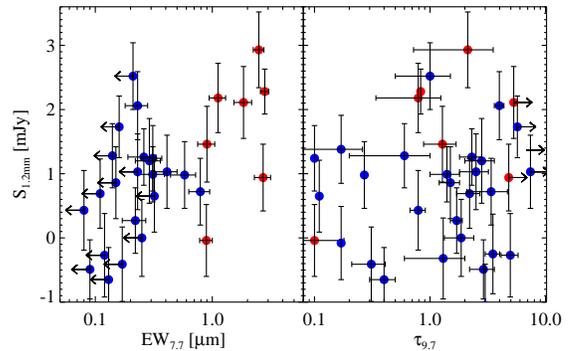}
\end{center}
\caption{{\it Left:} MAMBO\,1.2\,mm flux vs. EW7.7. To minimize the effects of negative $k$-correction, we only plot $z$\,$>$\,1.5 sources.  {\it Right:} MAMBO\,1.2\,mm flux vs. $\tau_{9.7}$.  We have added 0.1 to all $\tau$ values in order to be able to plot the sources with $\tau_{9.7}$\,=\,0.0. In all cases, the red symbols are strong-PAH sources (EW$_{7.7}$\,$>$\,0.8\um), while the blue symbols are weak-PAH sources (EW$_{7.7}$\,$<$\,0.8\um). \label{cold_origin}
}
\end{figure}

As shown in Table~\ref{table_det}, both the 160\um\ and 1.2\,mm detectability of our sample are rather low (21\% and 16\% respectively).  However, in both cases, the detectability is higher for the strong-PAH vs. weak-PAH $z$\,$>$\,1.5 sources (e.g. 50\% vs. 11\% for the 1.2\,mm data).  This is shown in Figure~\ref{cold_origin}{\it left}. To assess the strength of the perceived trend, we perform a Spearman rank correlation. We find $\rho$\,=0.569 with probability of correlation of 99.9\%.   By contrast, the MAMBO 1.2\,mm flux vs. the optical depth of the 9.7\um\ silicate feature (Figure~\ref{cold_origin}{\it right}) has a Spearman rank correlation of $\rho$\,=\,0.146 with probability of correlation of 68.2\%.  This suggests that there is a significant trend of strong-PAH sources to be strong cold-dust emitters, but no such trends exists with respect to the degree of obscuration as quantified by the depth of the silicate feature. This conclusion supports the finding of \citet{desai07} who show that in local ULIRGs, the degree of obscuration and the far-IR colors are not related. 

\subsection{Presence of AGN: $L_{\rm{radio}}$ and spectral indices \label{sec_lradio}}

\begin{figure}[!ht]
\begin{center}
\plotone{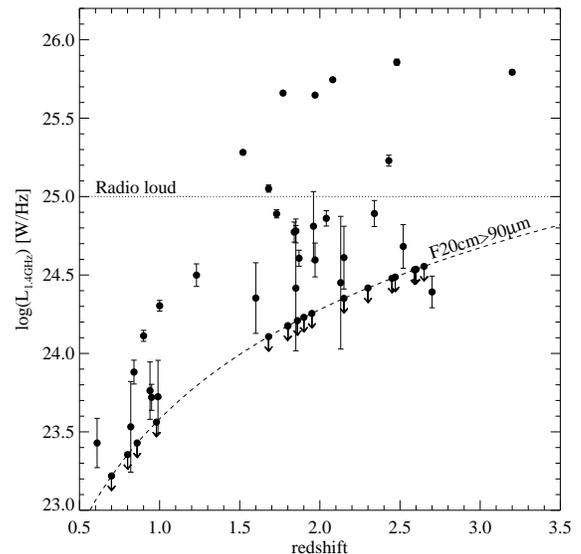}
\end{center}
\caption{Distribution of radio luminosities vs. redshift for our sample. The dotted line shows the radio-loud limit we adopt \citep{rl_letter}. The dashed line shows the XFLS 4\,$\sigma$ limit of 90\,$\mu$Jy (assuming $\alpha^{610\rm{MHz}}_{1.4\rm{GHz}}$\,=\,-0.7).  The upper limits are our radio-undetected sources which by definition lie on the dashed line. \label{lr_z}}
\end{figure}

Table~\ref{table_fits} gives the rest-frame 1.4\,GHz luminosities of our sample, where the $k$\,-\,correction is  based on their observed $\alpha^{610\rm{MHz}}_{1.4\rm{GHz}}$ spectral indices (see Table~\ref{table_long}),  or assuming $\alpha^{610\rm{MHz}}_{1.4\rm{GHz}}$\,=\,-0.7  for the sources not detected in 610\,MHz.  In \citet{rl_letter}, we adopted $L_{1.4\rm{GHz}}$\,$>$\,$10^{25}$W/Hz as our radio-loud criterion.  Figure~\ref{lr_z} shows the radio luminosities for the entire sample.  It is obvious that most of the non-radio loud, radio detected sources still have significant radio luminosities ($L_{1.4\rm{GHz}}$\,$\sim$\,$10^{24}$\,--\,$10^{25}$\,W/Hz).  Most sources in this range are likely AGN powered as well.  For example, the radio luminosities of SMGs, which were found to be extended (i.e. starburst-powered) in high-resolution radio imaging all have radio luminosities $<$\,10$^{24}$\,W/Hz \citep{chapman04}.  

The radio spectral index, $\alpha$, is typically assumed to be $\sim$\,-0.7 for star-forming galaxies (Condon 1992), and $\sim$\,-0.5 for quasars \citep{stern00}. Recent blank-sky radio observations are consistent this range \citep{bondi07,huynh07a}. Steep spectra ($\alpha$\,$\sim$\,-1) are however the norm for high-$z$ radio galaxies (HzRGs). Flat or inverted spectra (essentially $\alpha$\,$>$\,-0.4) are strongly indicative of AGN.

\begin{figure}[!ht]
\begin{center}
\plotone{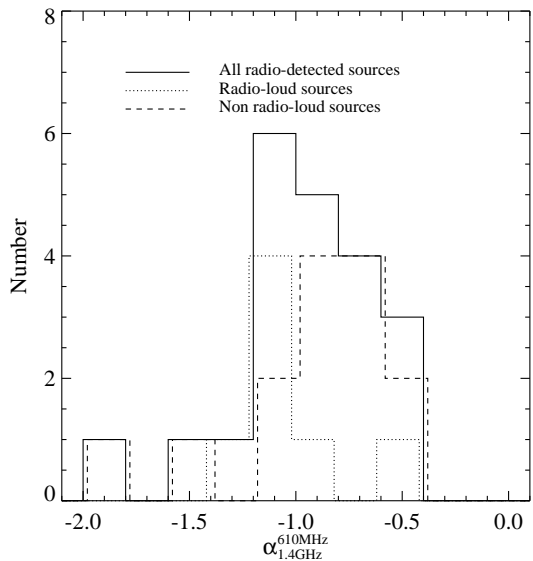}
\end{center}
\caption{Distribution of radio spectral index for the sources with both VLA 1.4\,GHz and GMRT 610\,MHz detections.  We also show the radio-loud and non-radio-loud sources separately.  \label{radio_alpha}}
\end{figure}

Figure~\ref{radio_alpha} shows the distribution of $\alpha^{610}_{1.4}$ for all sources with both VLA and GMRT detections, as well as the radio-loud and non-radio-loud sources separately.  The bulk of our radio-detected, but not radio-loud sources show fairly standard spectral indices ($\alpha^{610}_{1.4}$\,$\sim$\,-0.8). As discussed in \citet{rl_letter}, our radio-loud sample tend to have steep radio spectral indices ($\alpha^{610}_{1.4}$\,$\ls$\,-1).  This is consistent with the steep spectra found for a sample of $z$\,$\sim$\,2 Type2 quasar candidates \citep{alejo06b}. \citet{manuela08} also find that the bulk of their optically-faint 24\um-bright sources show steep spectra ($\alpha^{610}_{1.4}$\,$\ls$\,-1.0)).

Three of our sources (MIPS227, MIPS16059\footnote{This source shows an inverted spectrum, given its 1.4\,GHz flux of 0.57\,mJy and no 610\,MHz detection. The formal 610\,MHz detection limit (3\,$\sigma$ we adopt is 0.3\,mJy (see \S\,\ref{sec_gmrt}).  However, MIPS16059 is in a noisier part of the GMRT map and hence we estimate that a detection limit of $\sim$\,0.6\,mJy is more appropriate, which means the radio spectrum of MIPS16059 may also be flat.}, and MIPS22303) are consistent with flat spectra ($\alpha^{610\rm{MHz}}_{1.4\rm{GHz}}$\,$>$\,-0.4), by virtue of being undetected at 610\,MHz.  All of these sources are weak-PAH sources. For convenience, we have added a column in Table~\ref{table_fits} indicating whether or not a source shows signs of the likely presence of a radio AGN (i.e. is it radio-loud and/or has a flat spectrum).
Given that the typical uncertainty in the radio luminosity is $\sim$\,0.2\,dex, we also identified borderline radio-loud sources (i.e. those with $\log L_{\rm{1.4GHz}}$\,=\,24.8\,--\,25.0). As discussed above, these luminosities are also more than a factor of five higher than the strongest known starburst-powered radio luminosities. These are also very likely AGN-powered. 

We find that 12/34 (35\%) of our weak-PAH sources show signs of AGN.  Of the strong-PAH sources, 4/14 (29\%) show signs of radio AGN.  This shows that there is no clear trend between the PAH strength and radio-AGN detectability. However, in \citet{rl_letter}, we showed that of the high-$z$ weak-PAH sources, only sources with $\tau_{9.7}$\,$>$\,1 showed radio-loud luminosities (specifically 40\% of the sources in this category were radio-loud).  The reason for this finding is still unclear. 

\subsection{Presence of AGN: optical line diagnostics \label{sec_opt1}}

\subsubsection{High ionization lines \label{sec_highion}}

\begin{figure}[!ht]
\plotone{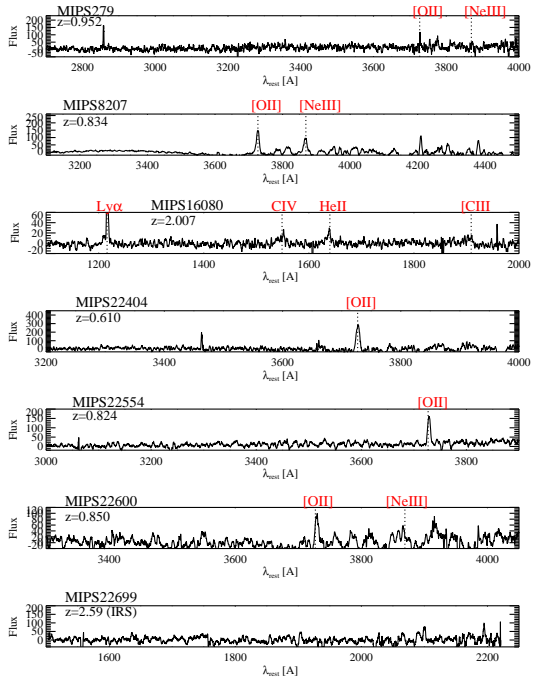}
\caption{The available rest-frame UV/optical spectra for our sample. Where detected, a continuum has been subtracted. The y-axis are in arbitrary units. The spectra for MIPS110, MIPS133, and MIPS15928 have been smoothed by a factor of 5 for clarity.\label{lris_all}}
\end{figure}

Figure~\ref{lris_all} shows the 13 available optical (rest-frame UV) spectra for our sample.  These include 12 Keck LRIS and DEIMOS spectra (\S\,\ref{sec_opt_specs}).  .  The rest-frame UV spectrum of MIPS22204\footnote{This source is the same as AMS13 in \citet{ams07}.} was presented in \citet{ams07}. For completeness, we include it here as well.

MIPS110 shows the [Ne{\sc v}]\,3426\AA\ line, which requires an AGN for excitation.  MIPS227, MIPS15928, MIPS16080, and MIPS22204 all show the C{\sc iv}1550\AA\ emission line. Our velocity resolution at 1550\AA is $\sim$\,280\,km/s.  In all cases, the line is resolved with a FWHM of $\sim$\,1241\,km/s, 2058\,km/s, 2507\,km/s, and 1305\,km/s respectively for MIPS227, MIPS15928, MIPS16080, and MIPS22204. These widths are consistent with the Narrow-Line Region (NLR) of AGN\citep{sm99}.  All these sources are therefore consistent with hosting obscured AGN. With the exception of MIPS15928 (which is a borderline strong-PAH source, see PaperII), all these sources are have weak-PAH mid-IR spectra. By contrast, the five sources which do not  show signs of AGN (MIPS283, MIPS8207, MIPS22404, MIPS22554, and MIPS22600) are all low-$z$, strong-PAH sources. The spectrum of MIPS133 is too red for the above diagnostic lines. However, we observe the [O{\sc iii}] line (see \S\,\ref{sec_lineratios}).  MIPS22699 shows no lines in its DEIMOS spectrum, which is not surprising if we assume its IRS-derived redshift of $z$\,=\,2.6.

{\plotone{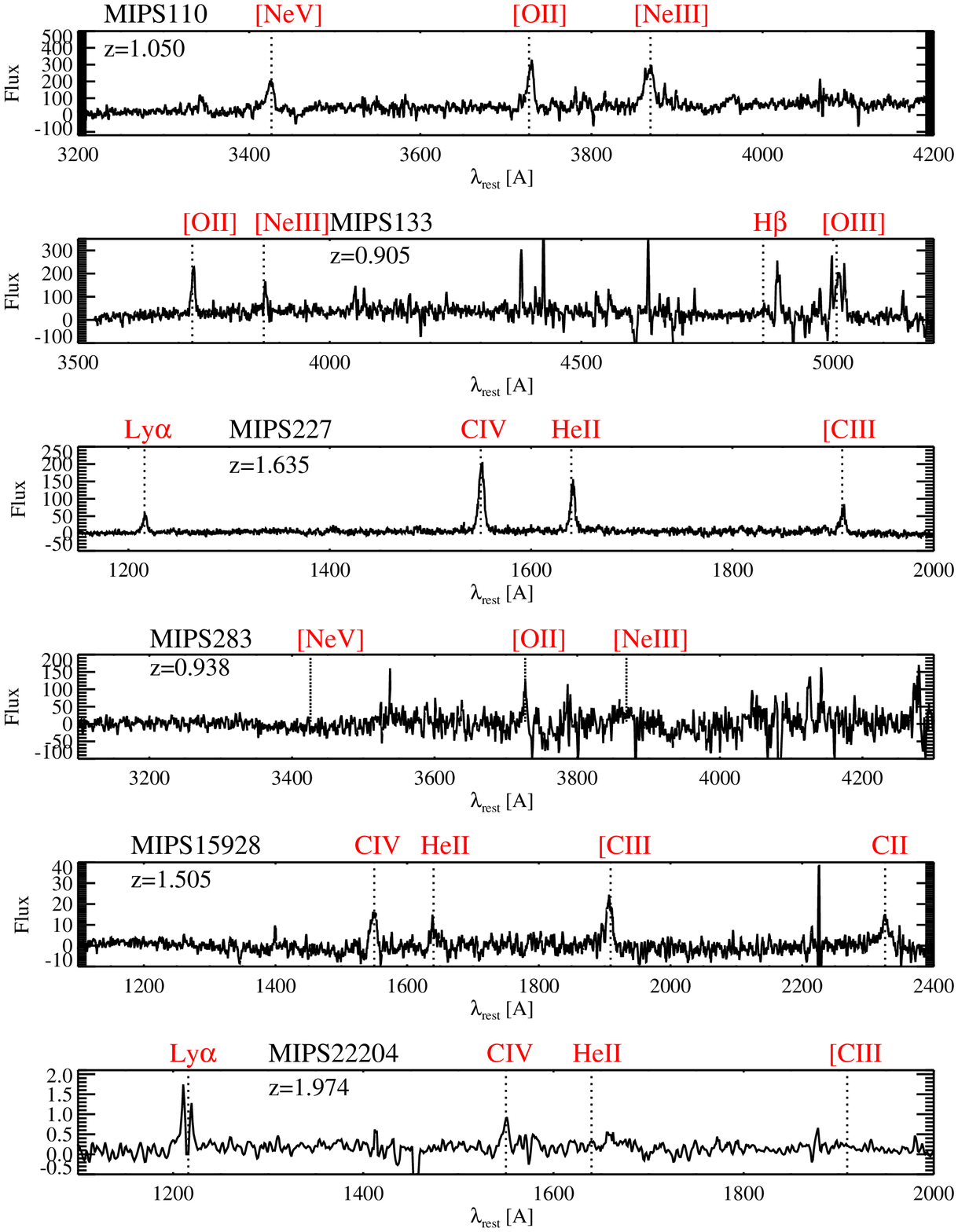}}
\centerline{Figure~\ref{lris_all} --- Continued.}

\subsubsection{Optical line ratios \label{sec_lineratios}}

\begin{figure}[!ht]
\begin{center}
\epsscale{.7}
\plotone{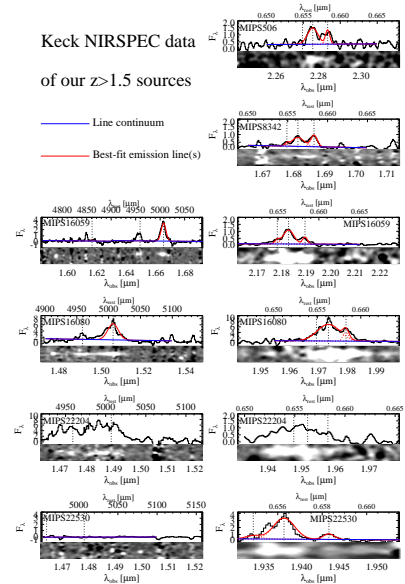}
\end{center}
\caption{The Keck NIRSPEC data. The y-axis units are $10^{-18}$\,erg/s/cm$^2$/\AA. We simultaneously fit the principle lines with Gaussian profiles of fixed central wavelength (see text for details). The best-fit linear continua are shown by the blue lines, while the red lines show the best-fit continuum+emission lines model.  The fit for MIPS22204 is omitted as it appears to be a blended source. For each source, the top panel shows the 1-D spectra, while the bottom shows the 2-D spectrum. The later are difference images (to subtract the sky lines) and are based on individual observations, not their mean as the 1-D spectra. The 2-D spectra have been smoothed with a 5 pixel kernel to highlight the features. \label{keck_all}}
\end{figure}

\begin{figure}[!ht]
\begin{center}
\epsscale{1}
\plotone{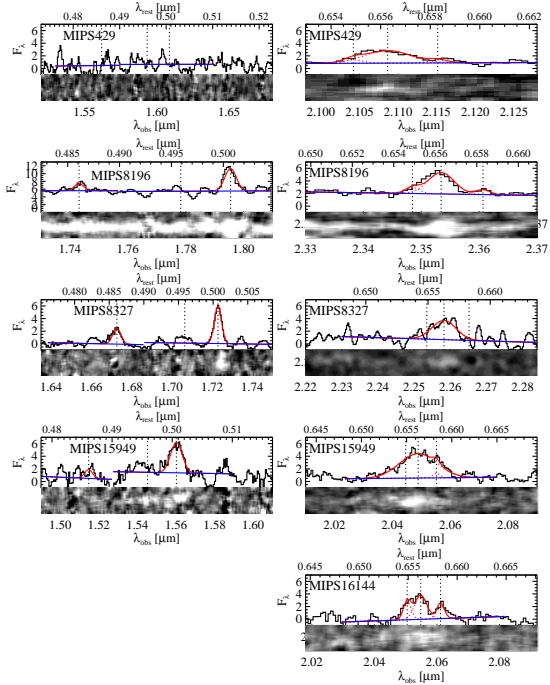}
\end{center}
\caption{The same as Figure~\ref{keck_all}, but for the Gemini NIRI spectra. The only exception is MIPS8327, whose $K$-band spectrum was taken with NIRSPEC. \label{gemini_all}}
\end{figure}

Figure~\ref{keck_all} shows the rest-frame optical spectra for the 6 sources\footnote{An additional source observed with NIRSPEC was MIPS78; however, no lines were detected in its spectrum. Its IRS-based redshift is $z$\,=2.65, which means that H$\alpha$ falls outside its observed range. It is also quite faint ($m_{H}$\,=\,21.98; Dasyra et al. 2008).}  with Keck NIRSPEC data and Figure~~\ref{gemini_all} shows the same for 5 sources with Gemini NIRI data. The H$\beta$, [O{\sc iii}]$\lambda\lambda$4959,5007\AA\ as well as the H$\alpha$6365\AA\ and [N{\sc ii}]$\lambda\lambda$6549,6583\AA\ complex are all marked.  These emission lines are fit with Gaussian profiles of fixed central wavelength but variable width and amplitude. The complex consisting of H$\alpha$ and the two [N{\sc ii}] lines is fit simultaneously with three Gaussians all of fixed central wavelength, and includes a single linear continuum component.  We determine the redshifts to fit the best both the $H$\,-\,band and $K$\,-\,band lines and estimate the near-IR redshift uncertainties at $\pm$\,0.001. The only exception is MIPS15949 where we found different redshifts were needed for the two (see \S\,\ref{sec_asymm}). In this fitting we use the {\sc idl} routine {\sc mpfit}. The principle line strengths and line ratios are given in Table~\ref{table_keck}.  The FWHM values have been corrected for the respective instrument resolutions (see \S\,\ref{sec_specs}). 

We adopt the following criteria for detecting an optical AGN:\\
1) FWHM(H$\alpha$)\,$>$\,2000\,km/s \citep[e.g.][and references therein]{zakamska03} \\
2) log([N{\sc ii}]/H$\alpha$)\,$>$\,-0.2. -- This is based on \citet{kauffmann03}, which cut likely includes composite starburst-AGN sources \citep[see e.g.][]{kewley06}.  This criterion is used only when log([O{\sc iii}]/H$\beta$) is not available.\\
3) When both ratios are available, we use the \citet{kewley06}  criteria to determine whether or not a source is a starburst, an AGN, or a composite source.

Table~\ref{table_keck} shows the classification of our sources according to the above criteria. We find that only 1 of our 11 sources (MIPS15949) can be classified as a broad-line (FWHM$>$\,2000km/s) AGN. Another potentially broad-line source is MIPS22204; however, we suspect it to be a blend of two sources nearly overlapping on the sky and with a small velocity separation ($\Delta v$\,$\sim$\,500\,km/s). The redshift ($z$\,=\,1.974) is based on the Ly$\alpha$ and C{\sc iv} detection presented in \citet{ams07}. Based on its C{\sc iv} line we classified it as an AGN in \S\,\ref{sec_highion}.

We find that, for the sources with optical/near-IR spectra, 3/9 (33\%) of the strong-PAH sources show signs of an optical AGN, while 9/13 (69\%) of the weak-PAH sources show signs of AGN. The much higher fraction of the later indicates that there is at least some correspondence between the mid-IR AGN diagnostics (i.e. PAH strength) and the optical AGN diagnostics. We cannot fully exclude the possibility that some of even our weak-PAH sources are pure starbursts with unusually strong mid-IR continua (e.g. MIPS429, MIPS8196); however, the bulk of the weak-PAH sources do show evidence of obscured AGN.  

\subsubsection{Asymmetric [O{\sc iii}] profiles \label{sec_asymm}}

\begin{figure}[!ht]
\begin{center}
\plotone{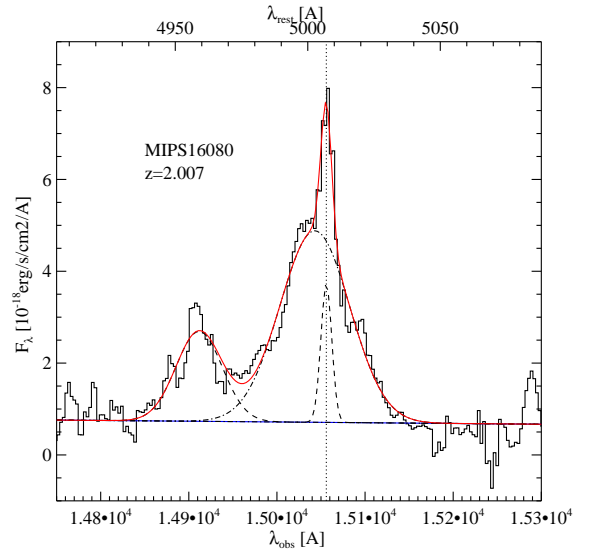}
\end{center}
\caption{The [OIII] profile of MIPS16080 showing a broader blueshifted component in addition to the narrow lines.  \label{fig_asymm}}
\end{figure}

An additional, diagnostic is the profile shape of the [OIII]\,5007\AA\ line.  Because of the poor signal to noise in many of the spectra, this is difficult to ascertain in most sources; however, some sources show clearly the blue-asymmetry frequently observed in the [OIII] lines of AGN \citep{heckman81,rice06}. This asymmetric profile is believed to be a combination of excess blue wings due to outflows. and obscuration of the red wings \citep{heckman81}.  Figure~\ref{fig_asymm} shows a close-up of the [O{\sc iii}] line of MIPS16080, which shows this asymmetry most clearly. We have performed a simultaneous fit of 3 Gaussians two of which are fixed to the [O{\sc iii}]\,5007\AA\ and 4959\AA\ lines, the central wavelength of the third line is not fixed but is restricted to be in between the other two. The best-fit additional broad component has FWHM\,=\,1849\,km/s and a velocity offset from the 5007\AA\ line of $\Delta v$\,=\,-261\,km/s.  

Another potential case of such an outflow is MIPS15949. We find that there the [O{\sc iii}] redshift is 2.116, while the H$\alpha$ redshift is 2.122. This blueshift of the [O{\sc iii}] line, if attributed to an outflow, implies outflow speed of 576\,km/s.  

\subsubsection{Emission-line based AGN luminosities}

Two effects make the determination of the intrinsic AGN $L_{\rm{[OIII]}}$ uncertain in our typically composite sources: the unknown level of star-formation contribution to this line, and the unknown level of extinction. Despite these caveats, the observed [O{\sc iii}] luminosities of our sources can still serve as a rough gauge of the strength of the central source. The [O{\sc iii}] luminosities are in the range  $\sim$\,10$^{8.56-9.02}$\lsun, or $\sim$\,10$^{42.2-42.6}$\,erg/s. These are comparable to the more luminous of the Type 2 quasar candidates selected in the SDSS by \citet{zakamska03}.  Without reddening correction on the emission lines, \citet{netzer06} derive a relationship between the [O{\sc iii}] and X-ray luminosities of Type2 quasars: log(L$_{\rm{[OIII]}}$ / L$_{2-10 keV}$)\,=\,(16.5\,$\pm$\,2.9) - (0.42\,$\pm$\,0.07) * L$_ {2-10 keV}$. Based on this, we expect intrinsic X-ray luminosities in the range $\sim$\,$10^{44.3-45.0}$erg/s for our sources with [O{\sc iii}] detections.  Even accounting for the flux calibration uncertainty in the spectra (see \S\,\ref{sec_calib}), our sources are consistent with quasar-strength AGN.  The {\sl Chandra} X-ray observations of our sample and obscuring columns are presented in Bauer et al. (2008, in prep.).  

\subsubsection{Optical Extinction and SFR \label{sec_optext}}

We have four sources with an estimate of the H$\alpha$/H$\beta$ ratio. These are: MIPS8196, MIPS8327, MIPS15949, and MIPS16059 with ratios of 2.5, 2.0, 9.8, and $>$\,25.7 respectively.  In a comprehensive study of 2000 Sy1 AGN, \citet{zhou06} find a mean value of H$\alpha$/H$\beta$\,=\,3.03, consistent with theoretical expectations, \citep[see][]{osterbrock}. The observed range in \citet{zhou06} is roughly H$\alpha$/H$\beta$\,$\sim$\,2\,--\,5. Our measured ratios may suffer some systematic uncertainty due to the uncertainties in the absolute flux calibration of the $H$ and $K$ bands (see \S\,\ref{sec_calib}). We can however conclude that MIPS8196 and MIPS8327 have low levels of optical extinction (if any), while MIPS15949 and MIPS16059 show significant extinction. 
Using a standard Milky Way extinction curve, we obtain E(B-V)\,=\,1.2 and E(B-V)\,$>$\,2.2 for MIPS15949 and MIPS16059 respectively. For comparison, \citet{brand07} find that all their IR-bright, optically-faint $z$\,$\sim$\,2 sources with H$\beta$ coverage, show strong extinction in the range E(B-V)\,=\,1.0-1.9.  

The two sources without optical extinction, MIPS8196 and MIPS8327, have respectively $\tau_{9.7}$\,=\,1.3 and 2.4 (PaperII). The two sources with strong optical extinction, MIPS15949 and MIPS16059, have respectively $\tau_{9.7}$\,=\,0.0 and 2.7. This suggest that there is no correspondence between the optical extinction values derived above and the strength of the 9.7\um\ silicate absorption feature. This is consistent with the conclusions of \citet{brand07}. In both cases, we are dealing with fairly small number statistics; however, this conclusion is not surprising given the expected complicated dust/power source geometries in these sources. 

The average $L_{H\alpha}$ of our sample (excluding the broad-line MIPS15949) is 2\,$\times$\,10$^{42}$\,erg/s (Table~\ref{table_keck}). This compares well with the non-broad line sources in \citet{brand07}. Based on \citet{ken98}, we derive $\langle$SFR(H$\alpha$)$\rangle$\,=\,9.8\msun/yr (range 1.3\,--\,42.4\msun/yr).  This is a lower limit given the expected extinction in most of these sources.  For MIPS8196 and MIPS8327 (two weak-PAH sources without significant optical extinction), we find SFRs of 22\msun/yr and 26\msun/yr respectively.  These modest SFRs are consistent with the far-IR non-detection of both these sources (see Table~\ref{table_long}).  We return to the question of IR-derived SFRs in the following section.  

\subsection{Far-IR SED fitting \label{sec_fit}}
In \S\,\ref{sec_origin}, we show that, especially in strong-PAH sources, but also in some weak-PAH sources, substantial cold dust emission is present further supporting the presence of starbursts in these sources. Here, we fit the available far-IR data (70\um, 160\um, and 1.2\,mm) in order to both determine the total infrared luminosities of our sources, and constrain the relative contribution of AGN and starburst processes to those luminosities.  
 The far-IR SED of infrared-luminous galaxies is frequently modeled as a composite of two or more different temperature modified blackbody components \citep[e.g.][]{dunne01, klaas01}.  The modification is given by $\nu^{\beta}$. The emissivity index, $\beta$, ranges between $\sim$\,1\,--\,2. The greybody approach is only an approximation to a much more complex radiative transfer problem, but is useful in characterizing far-IR SEDs when only a few data points are available. We only have data at 70\um, 160\um, and 1200\um, which, at $z$\,$\sim$\,2, correspond to $\lambda_{\rm{rest}}$\,$\sim$\,20, 50 and 400\um. 
 
\begin{figure}[!ht]
\begin{center}
\plotone{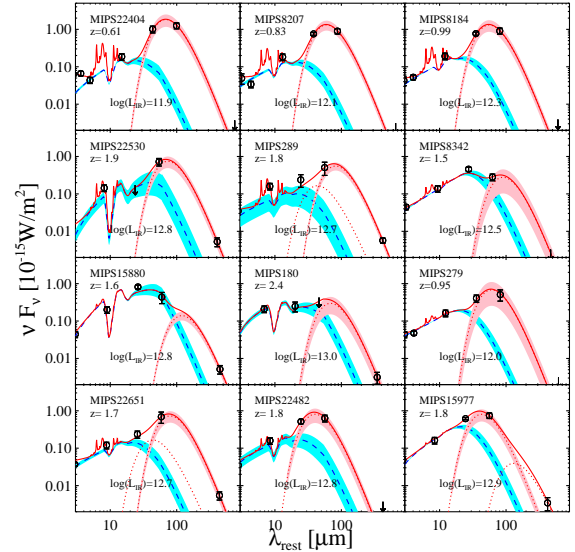}
\end{center}
\caption{The $\sim$\,1\,--\,1000\um\ SEDs of our sources with far-IR detections. The blue curve shows the best-fit to the mid-IR spectra (see PaperII). The dotted red curve shows the best-fit cold dust component with the pink shaded region being the 1\,$\sigma$ scatter.  In a few cases with all three far-IR points detected, a second greybody component was included to give a `smoother' SED, but without significantly contributing to $L_{\rm{IR}}$. For MIPS289 and MIPS22651 this component has $T_{\rm{d}}$\,=\,60\,K, while for MIPS15977 it has $T_{\rm{d}}$\,=\,20\,K. Upper limits are 2\,$\sigma$. In all cases, the solid red curve is the total SED.  \label{fullir_plots}}
\end{figure}

 In PaperII, we already discussed that our mid-IR continuum can be thought of as a sum of different temperature dust components ranging from $\sim$\,1500\,K to $\sim$\,100\,K.  Here, we assume that this mid-IR continuum is entirely AGN-powered, and use this model and set the coldest AGN component at $T_{\rm{outer}}$\,=\,100\,K, $\beta$\,=\,1.5. This choice gives good fits for the vast majority (83\%) of the sources; however, some exceptions exist. In particular, for MIPS506, MIPS8242, and MIPS15840 we used $T_{\rm{outer}}$\,=\,150\,K, while for MIPS8342, MIPS15928, MIPS16080, MIPS16095, and MIPS22277 we found that $T_{\rm{outer}}$\,=\,80\,K gave better agreement with the data. This continuum of temperatures was additionally subjected to the extinction of a level determined by the 9.7\um\ silicate feature depth (see PaperII). Our fitting procedure in PaperII includes the uncertainty in the determination of this mid-IR continuum which was especially strong for high-$z$, strong-PAH sources. We find however, that the uncertainty in the total $L_{\rm{AGN}}$ requires a further consideration of the uncertainty in $T_{\rm{outer}}$, which is the dominant component for steep-slope sources. This is difficult to quantify robustly, but we find that a range of $\pm$\,10\,K in its value roughly corresponds to the uncertainty in the MIPS70\um\ points (rest-frame $\sim$\,20\um). In Figure~\ref{fullir_plots}, we show the AGN component with its 1\,$\sigma$ uncertainty, which includes both the fitting uncertainty from PaperII and this reasonable spread in the value of $T_{\rm{outer}}$. We further discuss the AGN-powered mid-IR SEDs in \S\,\ref{sec_agnsed}. 
 
Before fitting for the starburst-powered cold dust emission, we subtract the AGN-component as described above from the MIPS70 and MIPS160\um\ points. Because of the typical far-IR non-detection of our sources, or detection of 1\,--\,2 points only, we cannot justify fitting more than a single greybody component (there are three exceptions here as described in Figure~\ref{fullir_plots}). We use the {\sc idl} routine {\sc mpfit} to fit for this cold dust component and determine its 1\,$\sigma$ spread. In the fit, the amplitude, and temperature are free parameters while we fix the emissivity index, $\beta$\,=\,2. Degeneracy between $T_{\rm{d}}$ and $\beta$ \citep[see e.g.][]{me06} means that this high $\beta$ value leads to cooler temperatures.  

Figure~\ref{fullir_plots}, shows the best-fit SEDs for all sources where at least 2 of the 3 photometric points are detections. For the rest, the far-IR is much less constrained. However, in the vast majority of these cases, the AGN-component dominates (and hence $L_{\rm{IR}}$ is determined reasonably well despite the far-IR non-detections). The average SED of the weak-PAH, far-IR non-detections is discussed in \S\,\ref{sec_agnsed}.  The integrated 3\,--\,1000\um\ luminosities ($L_{\rm{IR}}$), as well as the luminosities of the AGN and starburst components separately are presented in Table~\ref{table_fits}.  The starburst luminosities listed in Table~\ref{table_fits} are  the {\it sum} of the cold dust emission and the PAH emission, determined from our mid-IR fits in PaperII. 

\subsubsection{Star-formation rates}

Lastly, using the standard relation from \citet{ken98}, we derive the IR-based SFRs for the 160\um-detected sources and upper limits for the majority of the sources which are non-detections. The $z$\,$<$\,1.5, 160\um-detected sources are typically strong-PAH sources whose implied SFR is $\sim$\,100\,--\,300\msun/yr.  The $z$\,$>$\,1.5 160\um-detected sources typically have SFRs of $\sim$\,900\msun/yr.  We believe that the upper limits for our strong-PAH sources are likely close to their SFRs. For the weak-PAH sources, while the upper limits allow for significant SFRs (up to $\sim$\,800\msun/yr), in at least two cases we find more modest SFRs of $\sim$\,20\,--\,30\msun/yr from their H$\alpha$ luminosities and low optical extinction levels (see \S\,\ref{sec_optext}). This is likely to be roughly the lower limit on the SFRs of the other weak-PAH non 160\um-detected $z$\,$>$\,1.5 sources. 

MIPS22530 is the only one source with both 160\um\ detection and an H$\alpha$ luminosity. For this source, the H$\alpha$ derived SFR is 7.3\msun/yr, while its IR-based SFR is 990\,$\pm$\,400\msun/yr.  This implies an extinction of A(H$\alpha$)\,=\,5.3 (or 4.8 if we take the lower limit on SFR$_{\rm{IR}}$). However, as can be seen in Figure~\ref{keck_all}, the removal of sky lines in this source is somewhat suspect and we might be missing some of the H$\alpha$ flux.  Nevertheless, reconciling the two SFR estimates clearly requires extreme levels of obscuration. For comparison, \citet{brand07} find A(H$\alpha$)\,$>$\,2.4, 3.8, and 4.6 in three of their $z$\,$\sim$\,2 ULIRGs with limits on the H$\alpha$/H$\beta$ ratio. Therefore, although extreme, the required H$\alpha$ extinction is not unprecedented in similar populations.

\subsection{$L_{\rm{MIR}}$\,--\,to\,--\,$L_{\rm{IR}}$ \label{sec_lir}}

An important parameter derived from the spectral fitting is the integrated 3\,--\,1000\um\ infrared luminosity, $L_{\rm{IR}}$ (see Table~\ref{table_fits}).  We find that our $z$\,$\sim$\,1 sources have $L_{\rm{IR}}$\,$\sim$\,\lir{12}, while the bulk of the sample which lies at $z$\,$\sim$\,2, have $L_{\rm{IR}}$\,$\sim$\,7\,$\times$\,\lir{12}.  
The greater luminosities of our high-$z$ sources are the usual consequence of a flux-limited sample (see Paper\,I for details of the Malmquist bias of our sample). But how reliable are our $L_{\rm{IR}}$ values given that we fit the far-IR SED from only three points: 70\um, 160\um, and 1.2\,mm?  At $z$\,$\sim$\,1, the 160\um\ point samples near the peak of the SED distribution, while for weak-PAH sources without far-IR detections, the mid-IR continuum dominates the infrared luminosity in any case. For $z$\,$\sim$\,2 MIPS160\um-detected sources we are missing the important $\sim$\,100\um\ regime. In practice, our fitting procedure typically leads to rather cold dust temperatures which peak at $\sim$\,100\um\ (see Table~\ref{table_fits} and Figure~\ref{fullir_plots}).  We estimate that if we assume warmer SED, i.e. ones that peak at the rest-frame of the MIPS160\um\ observations, then the quoted $L_{\rm{IR}}$ will be overestimated by $\sim$\,0.1\,dex. This is comparable to our $L_{\rm{IR}}$ uncertainties (see Table~\ref{table_fits}).

It is well known that the mid-IR luminosity of star-forming galaxies is correlated with the total infrared luminosity, although the exact relation derived is somewhat affected by selection biases \citep{ce01,tak05,nb07}.  However, these relations are optimized for the typical star-forming galaxy ($L_{\rm{IR}}$\,$\sim$\,$10^{10}$\,--\,$10^{11}$\lsun). It is not obvious how well they perform at predicting the infrared luminosities of ULIRGs, especially those with significant AGN contribution. The availability of IRS spectra and far-IR-to-mm data allows us to test the conversion between mid-IR broadband observations and $L_{\rm{IR}}$ for such extreme sources.     
 
\begin{figure}[!ht]
\begin{center}
\plotone{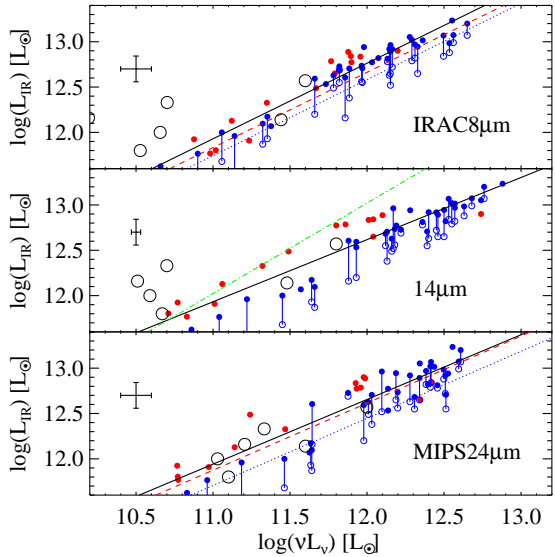}
\end{center}
\caption{{\sl Top panel:} The IRAC8\um\ luminosity vs. $L_{\rm{IR}}$. The solid line is our best-fit relation. We overplot the \citet{nb07} relations  based on their detected sources (red dotted line) and including the stacks of the non-detected sources (blue dotted line).  Whenever $L_{\rm{IR}}$ is an upper limit, we also give the expected range (open circle connected to a filled circle). {\sl Middle panel:} The monochromatic 14\um\ luminosity vs. $L_{\rm{IR}}$.  The best fit linear relation is shown by the solid line. The green dot-dash line is the 15\um\ conversion relation \citep{ce01}. We have ignored the difference between the {\sl ISO} 15\um\ luminosity and our monochromatic 14\um\ luminosity.  {\sl Bottom panel:} The MIPS24\um\ luminosity vs. $L_{\rm{IR}}$. The red dashed and blue dotted lines are as in the top panel. In all cases, the red filled symbols are strong-PAH sources, while the blue filled symbols are weak-PAH sources.  The open circles are local ULIRGs \citep{armus07}. \label{lum_conv}}
\end{figure}

In Figure~\ref{lum_conv}, we present the conversion between the monochromatic 14\um\ luminosity and $L_{\rm{IR}}$ as well as the {\it rest-frame} IRAC8 and MIPS24 luminosities for comparison with broadband data. Below we present the best-fit linear fits in the log-log plane for the 8\um, 14\um, and 24\um luminosities respectively:\\
\begin{eqnarray}
\log(L_{\rm{IR}})=(2.79\pm0.36)+(0.83\pm0.03)\log(\nu L_{\nu,8\mu\rm{m}})\\
\log(L_{\rm{IR}})=(4.43\pm0.43)+(0.68\pm0.04)\log(\nu L_{\nu,14\mu\rm{m}})  \\
\log(L_{\rm{IR}})=(4.13\pm0.54)+(0.71\pm0.04)\log(\nu L_{\nu,24\mu\rm{m}})
\end{eqnarray}

where the rms spread is respectively 0.11, 0.15, and 0.17\,dex for the three relations. We stress that the above are fits to our (extremely biased) sample. They are not meant to be all-purpose conversion relations, and should only be used for sources of comparable mid-IR luminosities to our sources.  Another major caveat is that for many of these sources, the infrared luminosities are upper limits only.  Figure~\ref{lum_conv} shows the expected range in $L_{\rm{IR}}$ for these sources. Accounting for the range in $L_{\rm{IR}}$, we find that the our sources roughly agree with the \citet{nb07} 8\um\ conversion relation.  However, our strong-PAH sources, with 160\um\ detections all have systematically higher $L_{\rm{IR}}$ per 8\um\ luminosity. For some local ULIRGs the gap is even more extreme. This result is not surprising given that lower luminosity starbursts have higher PAH contribution  to $L_{\rm{IR}}$ (from a few percent up to $\sim$\,20\% \citet{jdsmith06}), while PAH contribute from $\ls$1\% up to a few percent of the total IR luminosity of ULIRGs \citep{armus07}. 

For $L_{14}$ vs. $L_{\rm{IR}}$ (Figure~\ref{lum_conv}{\it middle}), we find the slope is significantly different from unity, implying a higher MIR-to-FIR ratio at the highest luminosities.  Local conversion relations for star-forming galaxies suggest a slope\,$\sim$\,1 \citep{ce01}.   As expected, our much shallower slope is dominated by the weak-PAH sources with their strong mid-IR continua. The strong-PAH sources, for the most part, agree reasonably well with the \citet{ce01} relation. 

Finally, Figure~\ref{lum_conv}{\it bottom} shows the rest-frame MIPS24\um, whose best-fit relation gives the best agreement with the lower luminosity conversion relation. The same caveats as discussed above for the IRAC8\um\ relation apply here as well. 

\subsection{The starburst contribution in weak-PAH sources \label{sec_agnsed}}

Apart from obtaining the total infrared luminosities of our sources, the SED fitting described in \S\,\ref{sec_fit} provides us with the relative contributions of  AGN and star-formation to $L_{\rm{IR}}$.  We wish to compare our empirical modeling with other approaches in order to assess the level of uncertainty in the $L_{\rm{AGN}}$ and/or $L_{\rm{SB}}$ values we obtain. In this section, we address this question for the bulk of our sample which consists of $z$\,$>$\,1.5, weak-PAH sources. We discuss the validity of our decomposition for strong-PAH sources in the following section (\S\,\ref{sec_sbsed}).  

\begin{figure}[!ht]
\begin{center}
\plotone{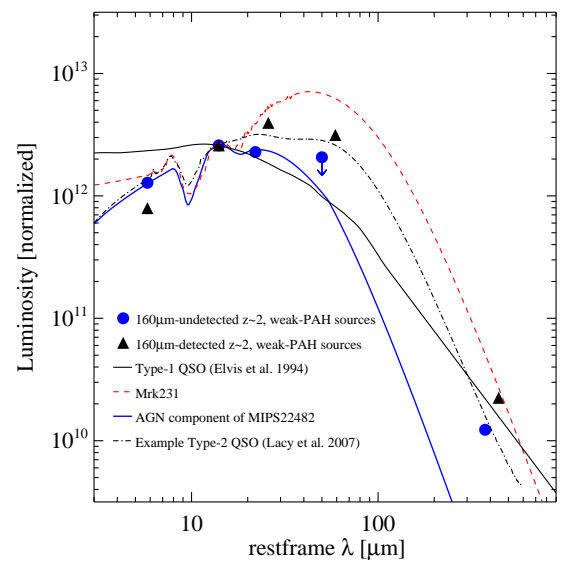}
\end{center}
\caption{The average SED of our 160\um-undetected $z$\,$\sim$\,2,weak-PAH sources, along with the AGN-component of one of our strong-PAH sources (MIPS22482). For comparison, we also show  the classic  \citet{elvis94} quasar SED, the Mrk231 spectrum and the SED of a Type 2 quasar from \citet{lacy07_letter}. All data and templates have been normalized at 14\um. \label{stacked_sed}}
\end{figure}

Figure~\ref{stacked_sed}, shows the average SED of the 160\um-undetected weak-PAH sources with $z$\,$>$\,1.5.  These are compared with the three 160\um-detected weak-PAH sources with $z$\,$>$\,1.5 (MIPS8342, MIPS15880, and MIPS15977). We overlay a number of templates including the classic QSO template of \citet{elvis94}, a torus model, the spectrum of the warm ULIRG Mrk231, and an example of a Type-2 AGN template \citep{lacy07_letter}.  We find that the 160\um-detected weak-PAH sources are roughly consistent with a Mrk231-like SED, but the stacked 160\um\ point of the non-detections (see above) is relatively weaker.  These are similar to the Type-2 quasar from \citet{lacy07_letter} (J1711+5953).  As is most likely the case with our sources, this source is also an AGN-starburst composite (note the presence of PAH features).  The presence of PAH features after stacking the weak-PAH sources (see PaperII), also suggests that their starburst fractions are $>$\,0.  However, we should remind that the stacked  MIPS160 is only an upper limit. Moreover, half of these sources are undetected at 70\um\  whose stacked value (1.9\,mJy) is about a factor of 2 below the faintest of the 70\um-detected sources. Therefore, we cannot exclude that some fraction of our sources are consistent with our adopted pure-AGN template. 

Our decomposition implies that the three weak-PAH, 160\um-detected sources have significant starburst components ($\sim$\,0.5\,$L_{\rm{IR}}$). By analogy, so does Mrk231, which has a similar SED. The composite (AGN+starburst) nature of Mrk231 is well established \citep{farrah03,armus07}.  Based on our AGN-template, Mrk231 has 63\% starburst fraction. By comparison, using the \citet{e_mrr} models, \citet{farrah03} derive a starburst fraction of 70\% for Mrk231.  More recently, \citet{fritz06} derive an total infrared starburst fraction of 73\% in Mrk231.  Our much simpler empirical approach comes within 10\% of these values. 

To further estimate the range of AGN luminosities from different models, we compare our values with the SED decomposition of \citet{polletta07}, which is based on a radiative transfer torus model and a starburst template. Polletta et al. derive $\log L_{\rm{AGN}}$\,=\,$\log L_{6\mu\rm{m}}$\,+\,0.32. For our sample the median ($\log L_{\rm{AGN}}$\,-\,$\log L_{5.8\mu\rm{m}}$) is 0.6 which makes $L_{\rm{AGN}}$ about a factor of 2 larger than if we had adopted the Polletta et al. conversion. However, much of this is due to our sample having particularly steep 24\um/8\um\ slopes. The Polletta et al. sample cover a much larger range in the 24\um/8\um\ colors. Their sample, however includes 5 of our $z$\,$\sim$\,2 sources (MIPS42, MIPS78, MIPS15840, MIPS22204, MIPS22303).  Converting their AGN luminosities to \lsun\ (and accounting for slight differences in redshifts), we find that $\log L_{\rm{AGN}}$ for these 5 sources is respectively: 12.91, 13.07, 12.71, 12.89, and 12.64.   Our values are always higher, but within $<$\,0.1\,dex of the Polletta et al. luminosities. The only exception is MIPS22303 where our $L_{\rm{AGN}}$ is 0.26\,dex higher. These examples suggest that our adopted empirical treatment of the AGN emission is roughly consistent with other approaches; however, the systematic uncertainty of individual sources' $L_{\rm{AGN}}$ can be as high as 0.3\,dex (a factor of 2).   

By extension, the `remainder' of the SED, which is attributed to starburst activity is also roughly consistent.  Therefore, we conclude that the starburst-fraction of weak-PAH sources ranges from $\ls$\,30\% for 160\um-undetected sources (e.g. MIPS78) to $\sim$60\% (e.g. MIPS15977) comparable to the prototypical warm ULIRG Mrk231.   

Accounting for the above systematic uncertainty, even the lowest $\log L_{\rm{AGN}}$ of our $z$\,$>$\,1.5 weak-PAH sources are all still $>$\,$10^{12}$\lsun. This means that the conclusion that our sources host quasar-strength AGN is independent of our particular SED fitting approach. 

\subsection{The AGN contribution in strong-PAH sources \label{sec_sbsed}}
For the quarter of our sample which show strong PAH emission features, and typically strong cold dust emission (\S\,\ref{sec_origin}), the question is reverse from the one in the previous section concerning the bulk of our sources which are weak-PAH emitters. Based on our spectral decomposition, the AGN fraction in the $z$\,$<$\,1.5 sources ranges from 13\% (MIPS22404) to 20\% (MIPS283), although this could be higher in half the sources due to their non-detection at 160\um.  These sources do not show signs of either radio or optical AGN.  Of the 8 $z$\,$>$\,1.5 strong-PAH sources, two (MIPS506 and MIPS15928) are actually of borderline PAH-strength (see PaperII). Not surprisingly therefore, we find that both show significant AGN fractions (50\% and 79\% respectively).  MIPS15928 also shows both a radio and an optical AGN (see Table~\ref{table_fits}).  In this section, we concentrate on the remaining 6 $z$\,$>$\,1.5 strong-PAH sources (MIPS289, MIPS8493, MIPS16144, MIPS22482, MIPS22530, and MIPS22651). Half of these show signs of a radio AGN (see Table~\ref{table_fits}). For these sources our derived AGN fractions range from 20\% (MIPS289) to 32\% (MIPS22482).  Based on our decomposition, the $z$\,$>$\,1.5 strong-PAH sources have $L_{\rm{AGN}}$\,$>$\,10$^{12}$\lsun\ (except MIPS8493, which has $L_{\rm{AGN}}$\,=\,11.8). These luminosities suggest our sources harbor obscured quasars. 

\begin{figure}[!ht]
\begin{center}
\plotone{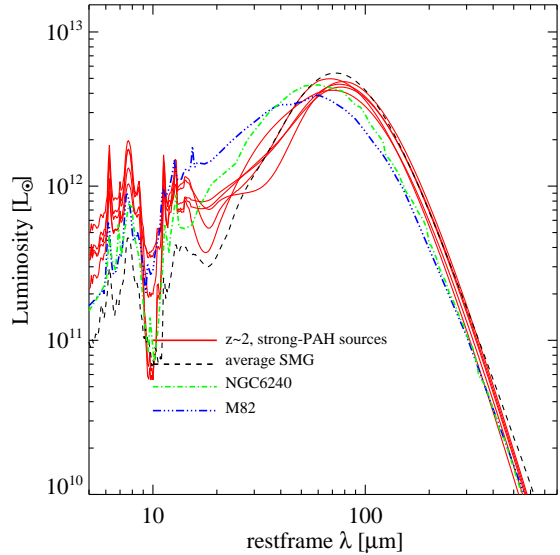}
\end{center}
\caption{The best-fit SEDs of our strong-PAH $z$\,$\sim$\,2 sources (see text for details).  These SEDs are compared with the average SED of SMGs \citep{pope06,pope07}, and the NGC6240 and M82 templates from the SWIRE template library (courtesy of M.Polletta, see also Weedman et al. 2006). Both our SEDs and the comparison templates are normalized so that their 3\,--\,1000\um\ luminosities are exactly 7\,$\times$\,$10^{12}$\lsun. \label{compare_all}}
\end{figure}

In our fitting, we have assumed that the $T$\,$\gs$\,100\,K continuum (peaking at $\sim$\,20\um) is due to an AGN. This leads to the 20\,--\,30\% AGN fractions in our $z$\,$\sim$\,2 strong-PAH sources (see Table~\ref{table_fits}).  However,  there is a wide range of relative mid-IR continuum levels even in pure starburst sources, emphasizing that, especially for strong-PAH sources, such infrared SED decomposition techniques are uncertain.  For comparison, in Figure~\ref{compare_all} we also show the SEDs of the pure starburst, M82, and of NGC6240 -- a starburst-dominated source but with significant AGN contribution \citep{risaliti06,armus07}. M82 has the stronger 10\,--\,30\um\ continuum, emphasizing the caution needed in assigning this continuum to an AGN.  It is however, worth recalling that M82 and sources like it are typically much lower luminosity, dwarf galaxies -- not obvious counterparts of our sources.  

The most closely related population to ours are SMGs which have the same redshifts and infrared luminosities are our sources. In Figure~\ref{compare_all}, we compare the spectra of these 6 $z$\,$\sim$\,2 strong-PAH sources with the average sub-mm galaxy \citep{pope06,pope07}.  By average sub-mm galaxy we mean, $z$\,$\sim$\,2, $S_{850}$\,$\geq$\,5\,mJy and $L_{\rm{IR}}$\,$\sim$\,$10^{13}$\lsun.  For simplicity, we exclude MIPS22482 from this figure because its best-fit greybody temperature is warmer (60\,K) than the rest ($\sim$\,30\,K). Its AGN fraction however is consistent with the three sources shown. All our SEDs have relatively stronger $<$\,30\um\ continua than the SMGs (especially MIPS22482).  The reality of this can be seen in that even though the redshifts and total infrared luminosities of our sources are quite similar to the SMGs, their observed 24\um\ fluxes ($>$\,0.9\,mJy) are consistently higher than those of SMGs \citep{pope07}.  Their observed 70\um\ fluxes ($\sim$\,5\,--\,12\,mJy) are also higher than the stacked 70\um\ flux of $z$\,$\sim$\,2 SMGs \citep[$\sim$\,2\,mJy; ][]{huynh07b}. Figure~\ref{compare_all}, shows that our strong-PAH sub-population is more similar to local ULIRGs such as NGC6240.  This extends our analysis of PaperII, where we conclude that the mid-IR properties of our sources are very similar to those of local ULIRGs, but at higher luminosities. Along with the SMGs themselves, this further supports the increasingly more clear conclusion that the Universe at  $z$\,$\sim$\,2 contained starburst-dominated systems of much higher luminosities than their local counterparts. 

\subsubsection{Dust mass}

We determine the dust mass only for the sources with MAMBO\,1.2\,mm detections (see Table~\ref{table_long}). These are most strong-PAH sources (see \S\,\ref{sec_sbsed}) as well as the three weak-PAH sources with strong cold dust component (MIPS8242, MIPS15880, and MIPS22558).  We use the standard method (see eg. Kovacs et al.~2006), which is given in Equation~\ref{eq-md}.

\begin{equation}
\label{eq-md}
M_{\rm{d}}= \frac{S_{\nu}\times D_{L}^2}{(1+z)\times\kappa_{\nu_{\rm{rest}}}\times B(\nu_{\rm{rest}},T)},
\end{equation}

where $T$ is the best-fit cold dust temperature from \S\,\ref{sec_fit}. The dust absorption efficiency, $\kappa$ is scaled from $\kappa_{850}$ using $\kappa_{\nu}\propto\nu^{-\beta}$. In the literature $\kappa_{850}$ varies between 0.08\,$\rm{m}^2\rm{kg}^{-1}$ and 0.3$\rm{m}^2\rm{kg}^{-1}$, depending on the dust properties (e.g. size, composition).  We use $\kappa_{850}$\,=\,0.15\,$\rm{m}^2\rm{kg}^{-1}$ as used by Kovacs et al. (2006) for a sample of sub-mm galaxies. We obtain dust masses of $\sim$\,5\,$\times$\,$10^8$\msun.  These dust masses are roughly comparable to those of SMGs \citep{kovacs06}. 

Assuming $M_{\rm{gas}}/M_{\rm{dust}}$\,$\sim$\,100 \citep{hildebrand83}, the above suggests $M_{\rm{gas}}$\,$\sim$\,$10^{10}$\,--\,$10^{11}$\msun. This is comparable to the gas reservoirs of SMGs. These have been shown to be sufficient to sustain the extreme star-formation rates needed to power their full infrared luminosisties ($\sim$\,$10^{13}$\lsun) \citep{tacconi06}.   Initial CO observations of some of our mm-bright sources support the speculation that these systems also contain large molecular gas masses (Tacconi et al. 2008, in prep.).

 \subsection{IR-radio correlation \label{ir_radio}}
 The far-infrared luminosity has long been known to correlate with the radio luminosity \citep{helou85}, the ratio of the two usually expressed by $q$.  The original definition of $q$ is based on the {\sl IRAS}\,60\um\ and 100\um\  fluxes, which samples the total 40\,--\,120\um\ power \citep{helou88}. Our data at $z$\,$\sim$\,2 spans the restframe $\sim$\,20\,--\,400\um, which allows us to measure directly {\sl IRAS}-equivalent luminosity and thus minimize $k$-correction effects. For consistency, we integrate our best-fit spectra to derive $L_{40-120}$. We rewrite the definition for the $q$ parameter to an equivalent, but more convenient form. 

\begin{equation}
\label{qform}
q=\log\left(\frac{L_{40-120}}{L_{\odot}}\right)-\log\left(\frac{L_{\rm{20cm}}}{\rm{WHz}^{-1}}\right)+14.03
\end{equation}
 
 To compute the rest-frame 20\,cm luminosity, we use the spectral indices, $\alpha$ (see \S\,~\ref{sec_lradio} and Table~\ref{table_long}).  For the sources without 610\,MHz detection, we assume the standard $\alpha$\,=\,-0.7.  
  
The relation remains linear over 4 orders of magnitude in luminosity, and for a wide variety of galaxy types with $\bar{q}$\,=\,2.34\,$\pm$\,0.01 and a standard deviation of 0.25\,dex \citep{yun01}. The origins of this tight relation are still unknown, although likely involve young stars fueling the infrared luminosity and, as supernovae, propelling the cosmic-ray electrons that fuel the radio luminosity \citep[see e.g.][]{murphy06}. Early results suggest that the relation holds at higher redshifts \citep{appleton04,frayer06}. Higher-$z$ can also mean more extreme luminosities are accessible. \citet{kovacs06} find that, for a sample of $z$\,$\sim$\,2 SMGs, the relation is lower than locally ($q$\,=\,2.07\,$\pm$\,0.09 using the standard $q$ definition).  
    
\begin{figure}[!ht]
\begin{center}
\plotone{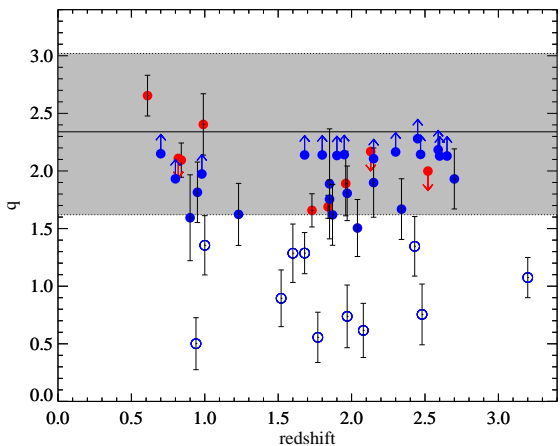}
\end{center}
\caption{The q-factor (far-IR to radio correlation) as a function of redshift. The solid line represents the local value, and the grey-shaded area is the 5\,$\sigma$ spread \citep{yun01}. As previously, the red symbols are strong-PAH sources. The blue symbols and the upper limits are weak-PAH sources.   
For clarity, we have marked the radio-loud sources with open symbols (they are excluded from the mean $q$ determination). \label{qplot}}
\end{figure}

Figure~\ref{qplot} shows the results for our sample. Excluding the radio-loud and radio-excess sources (i.e. $q$\,$<$\,1.6), we find $\langle q\rangle$\,=\,2.00\,$\pm$\,0.06 for all sources with radio detections (for lower and higher $z$). There is a weak trend with the $z$\,$\sim$\,1 (and $L_{\rm{IR}}$\,$\sim$\,\lir{12}) sources being closer to the local relation, while the $z$\,$\sim$\,2 (and $L_{\rm{IR}}$\,$\sim$\,\lir{13}) sources are consistently too low.  In general, steeper radio spectral indices lower $q$, while flatter spectra result in higher $q$.  To estimate the strength of this effect, we repeat the above, however assuming all sources without individual measurements of the spectral index have $\alpha$\,=\,-0.5.  This leads to $\langle q\rangle$\,=\,2.04\,$\pm$\,0.06. Therefore, the uncertainty in $\alpha$ does not significantly affect our conclusion of low $q$ values.  Another potential source of uncertainty is whether or not we have underestimated the infrared luminosities when fitting our sparsely sampled far-IR SEDs.  This issue was addressed in \S\,\ref{sec_lir} with the conclusion that this is probably not the case and indeed we are more likely to have {\it overestimated} the infrared luminosities. 

Figure~\ref{qplot} shows that roughly half the sources at $z$\,$>$\,1.6 (excluding the radio-loud sources) are undetected in the radio i.e. have only lower limits on $q$.  To estimate the effect of this population on the mean $q$ of our sample, we use the Kaplan-Meier estimator \citep{fn85}, which is non-parametric maximum likelihood estimator in the presence of censored data (i.e. lower or upper limits). Based on this we revise our estimate of for the $z$\,$>$\,1.5 sources to $q_{\rm{KM}}$\,=\,2.07\,$\pm$\,0.01, where $q_{\rm{KM}}$ denotes Kaplan-Meier estimated mean value of $q$.  

In order to answer the question whether the radio-undetected population itself obeys the local relation even if the sample as a whole does not, we also stack the radio images of the twelve sources without individual 1.4\,GHz detections. We find a mean flux of 57\,$\pm$\,6.6$\mu$Jy or an 8\,$\sigma$ detection.  Assuming this value for all twelve sources,  we find a mean $q$ of  2.21\,$\pm$\,0.02 which although still formally lower than the local mean value, is consistent with it given given the uncertain $\alpha$ values (see discussion above). Therefore, the radio undetected 37\% of the $z$\,$>$\,1.5 sources, as well as the $z$\,$<$\,1.5 sources are consistent with the local IR-radio correlation. 

A lower $q$ than the standard mean value is most commonly but not necessarily associated with excess radio emission due to an AGN. However, purely star-forming galaxies can exhibit infrared-to-radio ratios a few times lower than typical ($q$\,$\sim$\,2.0) in certain environments such as the centers of clusters \citep{mo01}, likely as a result of the thermal pressure of the Inter Cluster Medium. In rare cases,  a merger/interaction can form a synchrotron bridge between the merging pair which leads to lower $q$ values as well \citep{condon02}. When considering as yet poorly understood high redshift populations, such effects cannot be ruled out. 

\section{Discussion \label{sec_context}}

\subsection{A population of obscured quasars at $z$\,$\sim$\,2}

The AGN luminosities we derive based on our spectral decomposition ($\langle L_{\rm{AGN}}\rangle$\,=\,$10^{12.6}$\lsun) place our sample in the quasar regime. When available, the [O{\sc iii}] luminosities of our sources, imply intrinsic $L_{X-ray}$\,$\gs$\,10$^{44}$erg/s, supporting this conclusion.  As discussed in \S\,\ref{sec_agnsed} and \S\,\ref{sec_sbsed}, our SED decomposition (especially for strong-PAH sources) is uncertain. In particular, starburst-powered mid-IR continua have been observed both in conjunction with strong PAH (e.g. M82) and weak-PAH, but strong Si features (e.g. NGC1377; Roussel et al. 2006). The presence of either an optical or a radio AGN, however is independently supported for at least 20/35 (57\%) of our $z$\,$>$\,1.5 sample (see Table~\ref{table_fits}). These include both strong-PAH and weak-PAH sources.   Although, this does not prove the AGN-origin of their mid-IR continua, it  does make it the most likely explanation. We therefore, conclude that {\it at least} 57\% of our higher-$z$ sample are consistent with hosting obscured quasars, although this fraction can be as high as 100\%. The optical faintness, red infrared colors and frequent deep silicate absorption all suggests high degrees of obscuration. 

Two sources with silicate absorption do not show optical extinction in their H$\alpha$/H$\beta$ ratios suggesting perhaps an obscured nucleus surrounded by an unobscured host galaxy.  In \citet{rl_letter}, we argued that (the radio-loud) of these sources might be transition objects just before feedback effects dispel the obscuring matter and reveal the naked quasar. Some tentative evidence of such processes are seen in the blue-shifted [O{\sc iii}] emission of MIPS15949 and MIPS16080, which may indicate outflows of $\Delta v$\,=\,-576\,km/s and $\Delta v$\,=\,-261\,km/s respectively. 

We find that the number density of our $z$\,$\sim$\,2 sources is $\sim$\,2\,$\times$\,$10^{-6}$\,Mpc$^{-3}$. Their mean AGN luminosity is $\langle \log(L_{\rm{AGN}})\rangle$\,=\,12.6\,$\pm$\,0.2. Since these sources are optically faint, this luminosity is a close approximation to the bolometric luminosity.  The number density of unobscured (optically and X-ray selected) quasars of comparable bolometric luminosities at $z$\,$\sim$\,2 is a few\,$\times$\,$10^{-6}$\,Mpc$^{-3}$ \citep{hopkins07}. Therefore our sources are already comparable in number to the unobscured quasars of the same luminosity and redshift. The mean redshift of our sample, $z$\,=\,2.05, is also consistent with the epoch of peak quasar number density which is estimated at $z$\,$\sim$\,2.15 \citep{hopkins07}. Since our sample is not a complete census of the unobscured quasars at $z$\,$\sim$\,2 \citep[see also][]{polletta07} our results support the finding that obscured quasars at $z$\,$\sim$\,2 are at least as numerous, but likely outnumber the unobscured ones. 

Lastly, our sample is most likely characterized by sources powered by a combination of starburst and AGN activity (see \S\,\ref{sec_agnsed} and \S\,\ref{sec_sbsed}). Concurrent with the AGNs are, frequently ULIRG-strength starbursts accounting for $\sim$\,30\,--80\% of the bolometric emission.  These support the conclusion that a significant fraction of these obscured quasars are accompanied by vigorous star-formation activity \citep[see also][]{ams07}. 

\subsection{Implications for the SMG population}

We have shown that 6 of our 35 $z$\,$>$\,1.6 sources (20\%) have $S_{\rm{1.2mm}}$\,$>$\,2\,mJy, implying $S_{850\mu\rm{m}}$\,$\gs$\,5\,mJy.  These sources therefore would be classified as SMGs in blank-sky sub-mm surveys. Their number density however ($\sim$\,2/sq.deg.) is only $\sim$\,2\% that of the blank-sky sub-mm population \citep{coppin06,pope07}.  Four of these six sources are strong-PAH of similar PAH strengths as SMGs \citep{pope07}.  The other 2 of these strong mm sources, however, have continuum-dominated mid-IR spectra (MIPS8242 and MIPS15880).  In all cases, the mid-IR continua of our sources are found to be a higher fraction of their total infrared luminosities than is the case in SMGs (see \S\,\ref{sec_sbsed}). Our model suggests that this translates into higher AGN fractions. However, as our model assumptions here are uncertain, it is more accurate to say that these are upper limits on the AGN fractions.  Our spectral decomposition suggests $\sim$\,20\,--\,30\% AGN fractions of $L_{\rm{IR}}$ for the strong-PAH sources, while the weak-PAH sources have 40\,--\,79\% AGN fractions.  \citet{alexander05} find that the bulk (75\%) of SMGs host AGN; however, even after absorption-correction the typical $L_{\rm{X-ray}}$/$L_{\rm{IR}}$ for those SMGs is about an order of magnitude lower than that of quasars. This implies that their AGN contribute to only about 10\% of $L_{\rm{IR}}$.  This is roughly consistent with the estimates of Pope et al. (2007) based on mid-IR spectral decomposition. It appears likely that our few mm-bright sources represent the stronger-AGN tail of the SMG population. Particularly intriguing are the mm-bright sources with continuum-dominated spectra, because they are true composite objects where AGN and star-formation are inferred to each contribute roughly 50\% to $L_{\rm{IR}}$. These types of sources are still rare in our extragalactic samples, yet are clearly crucial in understanding any evolutionary links between starburst-dominated and AGN-dominated ULIRGs/obscured quasars. Such sources are likely to be detected in larger numbers in upcoming wide-field sub-mm surveys such as expected from SCUBA2 \citep{scuba2}, when combined with large-area mid-IR surveys such as SWIRE \citep{lonsdale03}. 

\section{Conclusions \label{sec_conc}}

1) Our sample shows low detectability in the far-IR (23\% at 160\um, and 16\% at 1.2\,mm). For the $z$\,$>$\,1.5 bulk of this sample, there is a trend of increasing MAMBO\,1.2\,mm fluxes with increasing PAH equivalent widths, but no clear trend with increasing 9.7\um\ silicate absorption.\\ 

2) By contrast 67\% of the sample are radio-detected. Based on their radio luminosities and/or spectral indices, we conclude the 17/35 (48\%) of our $z$\,$>$\,1.5 sources host radio AGN (about half of these are radio-loud).  This is supported by the observations of an enhanced radio emission for a given $L_{\rm{IR}}$ compared with expectations of the local IR-radio correlation. Excluding radio-loud sources, and accounting for the radio-undetected sources (through survival analysis), we estimate $\langle q\rangle$\,=\,2.07\,$\pm$\,0.01, which is significantly below the mean value ($q$\,=\,2.34) estimated locally. This radio excess is most likely, but not necessarily, due to the presence of radio AGN. \\

3) We further test for the presence of AGN using optical/near-IR spectroscopy which is available of 21 of the sources. We find a good correspondence between the mid-IR and optical diagnostics in that only 10\% of the strong-PAH sources show signs of optical AGN versus 62\% of the weak-PAH sources with optical/near-IR spectra. This supports on average the use of PAH-strength as a discriminant of AGN-dominated vs. starburst-dominated sources. However, 5 of our weak-PAH sources show starburst-like optical spectra.  Two of these show signs of radio AGN. \\

4) Of the 11 sources with near-IR spectra, we find only one broad line source \\(FWHM(H$\alpha$)\,$>$\,2000\,km/s). This is in contrast with \citet{brand07} who find that 7/10 of their high-$z$ obscured ULIRGs show broad emission lines.  On the other hand, the optical spectra of four of our sources show the C{\sc iv} emission line with widths corresponding to the NLR.  \\

5) Evidence for blueshifted [O{\sc iii}] emission is seen in two of our $z$\,$\sim$\,2 weak-PAH sources.  This is potentially indicative of outflows of up to 570\,km/s.\\

6) Where available, we find no correspondence between optical extinction and the depth of the silicate absorption feature.  We estimate that the lower limit on the star-formation rates in $z$\,$\sim$\,2 weak-PAH sources without far-IR detections is 20\,--\,30\msun/yr. By contrast, far-IR detected, $z$\,$\sim$\,2 sources (typically strong-PAH, see below) have SFR\,$\sim$\,900\msun/yr. \\ 

7)We fit the available far-IR data with a simple greybody model. We combine this with our mid-IR fitting, in order to estimate the total starburst luminosity (cold dust plus PAH), and AGN luminosity (hot/warm dust continuum). In sources with high PAH equivalent widths roughly $\sim$\,70\,--\,80\% of their total infrared luminosity is due to starburst activity. In weak-PAH sources this is typically $<$\,30\%. More intriguing are a a few weak-PAH sources with far-IR detections, suggesting significant starburst contributions ($\sim$\,50\%).  \\

8) We confirm expectations, that our sources have bolometric infrared luminosities of $\sim$\,$10^{12}$\lsun\ for the $z$\,$\sim$\,1 sub-sample, and close to $\sim$\,$10^{13}$\lsun\ for the $z$\,$\sim$\,2 majority of our sample.  We present relations between $L_{8\mu m}$, $L_{14\mu m}$ and $L_{24\mu m}$ and $L_{\rm{IR}}$.  \\

9) Compared with SMGs, our strong-PAH $z$\,$\sim$\,2 sources typically have similar redshifts, total luminosities, sub-mm emission (and hence dust mass), but higher mid-IR continua (attributed to greater AGN fractions). Our SED fitting suggests these sources have $\sim$\,20\,--\,30\% AGN contribution to their $L_{\rm{IR}}$. The presence of AGN is supported in 3/6 of these sources which have AGN-like radio luminosities. For our $z$\,$\sim$\,2, strong-PAH, MAMBO-detected sources, we find dust masses of $\sim$\,5\,$\times$\,$10^8$\msun, consistent with the values for sub-mm galaxies.  This suggests that these sources have comparable gas reservoirs to SMGs and hence are capable of sustaining the SFR$\sim$\,900\msun/yr implied by their AGN-subtracted infrared luminosities.\\

10) Finally , the presence of an AGN is confirmed through radio or optical line diagnostics in 57\% of the 35 $z$\,$>$\,1.5 sources.   This is a lower limit on the presence of AGN, because most AGN are not radio-loud, less than half our sources have optical/near-IR spectra and due to the high obscurations and composite nature of our sources an AGN may not be obvious in our optical spectra.  The infrared, radio, and optical spectroscopic diagnostics all support the conclusion that AGN contribute in the majority of these sources; however, predominantly we have composite starburst-AGN systems.  The AGN luminosities (based on the SED decomposition and [O{\sc iii}] luminosities where available), combined with the narrow emission lines and silicate absorption features suggest these sources host obscured quasars.  

\acknowledgements

We are grateful to the anonymous referee for their careful reading of our manuscript and detailed comments which greatly increased the clarity and presentation of this paper. We wish to thank Mark Lacy, Eric Murphy, Lee Armus, and Bruce Partridge for useful discussions.  We are grateful to Alexandra Pope for providing us with the average SMG SED template, and to Alejo Martinez-Sansigre for providing us with the optical spectrum of MIPS22204. We are grateful to D. Stern and G.Becker for allowing us to use their software tools to reduce the Keck spectra. This work is based on observations made with the {\sl Spitzer} Space Telescope, which is operated by the Jet Propulsion Laboratory, California Institute of Technology under contract with NASA. Support for this work was provided by NASA through an award issued by JPL/Caltech. This work includes observations made with IRAM which is supported by INSU/CNRS (France), MPG. (Germany) and IGN (Spain).  Some of the data presented herein were obtained at the W.M. Keck Observatory, which is operated as a scientific partnership among the California Institute of Technology, the University of California and the National Aeronautics and Space Administration. The Observatory was made possible by the generous financial support of the W.M. Keck Foundation. Lastly, this work includes observations obtained at the Gemini Observatory, which is operated by the Association of Universities for Research in Astronomy, Inc., under a cooperative agreement with the NSF on behalf of the Gemini partnership: the National Science Foundation (United States), the Science and Technology Facilities Council (United Kingdom), the National Research Council (Canada), CONICYT (Chile), the Australian Research Council (Australia), CNPq (Brazil) and SECYT (Argentina)


\clearpage

\begin{deluxetable}{lccccccc}
\tablecolumns{8}
\tablewidth{0pc}
\tabletypesize{\scriptsize}
\tablecaption{\label{table_long} Far-IR and radio fluxes with 1\,$\sigma$ uncertainties. }
\tablehead{\colhead{MIPSID} & \colhead{z} & \colhead{$S_{70}$\tablenotemark{a}} & \colhead{$S_{160}$\tablenotemark{a}} & \colhead{$S_{\rm{1.2mm}}$} & \colhead{$S_{\rm{20cm}}$} & \colhead{$S_{\rm{610MHz}}$}  & \colhead{$\alpha^{610\rm{MHz}}_{1.4\rm{GHz}}$}   \\
\colhead{} &  \colhead{} & \colhead{mJy} & \colhead{mJy} & \colhead{mJy} & \colhead{mJy} & \colhead{mJy} & \colhead{} }
 \startdata
 strong-PAH & & & & & &  \\
\hline 
  283 & 0.938\tablenotemark{c} & 6.4$\pm$2.2(2.8\arcsec) &  --- &  --- &  0.16$\pm$ 0.02  & 0.35$\pm$0.09 & -0.96 \\
  289 & 1.86 &  5.6\,$\pm$\,2.1(3.6\arcsec)\tablenotemark{b} &  54$\pm$10(20.0\arcsec)\tablenotemark{b} &  2.28$\pm$0.35 &  ---  & --- & --- \\
  506 & 2.470\tablenotemark{c} &  --- & --- &   1.37$\pm$ 0.53 &  0.14$\pm$ 0.03  & --- & $>$\,-0.92 \\
     8184 & 0.99 & 17.9$\pm$1.8(3.0\arcsec) &  49$\pm$9(5.0\arcsec)\tablenotemark{b} &  0.72$\pm$0.62 &  0.16$\pm$ 0.02  & 0.23$\pm$0.05 & -0.43 \\
      8207 & 0.834\tablenotemark{c} & 17.7$\pm$1.5(0.7\arcsec) &  48$\pm$8(4.8\arcsec) &   --- &  0.28$\pm$ 0.03  & 0.50$\pm$0.06 & -0.72 \\
 8493 & 1.80 &  --- & --- &  0.94$\pm$0.52 &  --- & 0.18$\pm$0.05 & $<$\,-0.83 \\
15928 & 1.505\tablenotemark{c} & 12.4$\pm$1.9(1.2\arcsec) & --- &  -0.04$\pm$0.56 &  1.25$\pm$ 0.06 & 2.93$\pm$0.15 &  -1.02 \\
16144 & 2.132\tablenotemark{c} &  --- &  --- &   2.93$\pm$0.59 &  0.12$\pm$ 0.03 & --- & $>$\,-1.10 \\
22404 & 0.610\tablenotemark{c} & 47.1$\pm$4.8(3.1\arcsec)\tablenotemark{b} &  134$\pm$12(9.4\arcsec)\tablenotemark{b} &  0.66$\pm$ 0.59 &  0.20$\pm$ 0.03 & --- & $>$\,-0.49 \\
22482 & 1.84 & 12.0\,$\pm$\,1.5(0.1\arcsec) &  34$\pm$7(1.5\arcsec) &   1.46$\pm$0.59 &  0.25$\pm$ 0.02  & 0.61$\pm$0.09 & -1.05 \\
22530 & 1.952\tablenotemark{c} &  --- & 38$\pm$9(9.0\arcsec) &  2.11$\pm$0.56 &  0.15$\pm$ 0.03  & 0.49$\pm$0.07 & -1.45 \\
22554 & 0.824\tablenotemark{c} & 10.2\,$\pm$\,1.7(2.2\arcsec) & --- &  --- &  0.13$\pm$ 0.03  & --- & $>$\,-1.01 \\
22600 & 0.850\tablenotemark{c} & 6.4$\pm$1.7(6.9\arcsec) & --- &  --- &  ---  & --- & --- \\
22651 & 1.73 &  5.5\,$\pm$\,1.2(3.8\arcsec) & 37$\pm$12(9.8\arcsec) &  2.18$\pm$0.54 &  0.47$\pm$ 0.03 & 0.94$\pm$0.09 &  -0.83 \\
 \hline   \\
 weak-PAH & & & & & &  \\
\hline 
   42 & 1.95 & 10.7$\pm$1.3(3.0\arcsec) &  --- & 0.43$\pm$0.62 &  --- & ---  & --- \\
    78 & 2.65 &  --- &  --- & -0.25$\pm$ 0.62 &  --- & --- & --- \\
  110 & 1.050\tablenotemark{c} &  9.5$\pm$2.2(7.2\arcsec)\tablenotemark{b} &  --- & -0.42$\pm$0.55 &  0.37$\pm$ 0.03  & 0.81$\pm$0.16 & -0.94 \\
  133 & 0.905\tablenotemark{c} & 12.3$\pm$2.0(1.3\arcsec) & --- &  0.01$\pm$0.57 &  0.39$\pm$ 0.03 & 0.73$\pm$0.09 & -0.77 \\
  180 & 2.47 &  5.8$\pm$1.7(4.3\arcsec) & --- &  1.26$\pm$0.44 &  ---  & -- & --- \\
 227 & 1.635\tablenotemark{c} & 10.1$\pm$2.0(3.4\arcsec) & --- &  -0.41$\pm$0.58 &  0.30$\pm$ 0.03  & --- & $>$\,-0.00 \\
 279 & 0.952\tablenotemark{c} &  9.5$\pm$2.1(1.0\arcsec)\tablenotemark{b} & 27$\pm$9(11.7\arcsec)\tablenotemark{b} &  0.31$\pm$ 0.50 &  0.35$\pm$ 0.06  & 0.86$\pm$0.19 & -1.10 \\      
  429 & 2.213\tablenotemark{c} &  --- & --- &  1.03$\pm$0.57 &  ---  & --- & --- \\
 464 & 1.85 &  7.3$\pm$1.5(3.2\arcsec) & --- &   -0.27$\pm$0.65 &  0.15$\pm$ 0.02 & --- & $>$\,-0.86 \\
  8034 & 0.95 &  --- &  --- &   0.74$\pm$0.62 &  0.14$\pm$ 0.02  & --- & $>$\,-0.92 \\
  8196 & 2.586\tablenotemark{c} &  5.0$\pm$1.4(5.6\arcsec)\tablenotemark{b} &  --- &  0.99$\pm$0.43 &  ---  & --- & --- \\
  8242 & 2.45 &  4.8$\pm$1.2(1.3\arcsec) &  --- &  2.52$\pm$0.52 &  --- & --- & ---  \\
 8245 & 2.70 &  --- & --- &  -0.49$\pm$0.46 &  0.130$\pm$ 0.009  & 0.24$\pm$0.06 & -0.74 \\
 8268 & 0.80 &  --- &  --- &  -0.45$\pm$0.78 &  0.047$\pm$\,0.008  & --- & --- \\
 8327 & 2.445\tablenotemark{c} &  --- & --- &  1.03$\pm$0.59 &  1.40$\pm$ 0.06  & 3.44$\pm$0.14 & -1.08 \\
 8342 & 1.562\tablenotemark{c} & 10.7$\pm$1.4(0.9\arcsec) & 29$\pm$11(8.8\arcsec)\tablenotemark{b} &  0.98$\pm$0.52 &  0.18$\pm$ 0.03 & 0.86$\pm$0.11 & -1.90 \\
 15840 & 2.30 &  --- & --- &  -0.08$\pm$0.57 &  ---  & --- & --- \\
 15880 & 1.68 & 19.0$\pm$2.1(1.1\arcsec) & 23$\pm$8(9.3\arcsec) &   2.06$\pm$0.53 &  0.60$\pm$ 0.21 & 1.35$\pm$0.06 &  -0.98 \\
15949 & 2.118\tablenotemark{c} &  7.3$\pm$1.6(0.6\arcsec) & ---- &   1.24$\pm$0.51 &  0.16$\pm$ 0.02  & --- & $>$\,-0.76 \\
15958 & 1.97 &  --- & --- &   1.28$\pm$0.50 &  0.27$\pm$ 0.02  & 0.39$\pm$0.07 & -0.45 \\
15977 & 1.85 & 14.3$\pm$1.0(1.1\arcsec) & 40$\pm$7(3.5\arcsec) &   1.38$\pm$0.53 &  0.298$\pm$ 0.008 & 0.61$\pm$0.07 &  -0.95 \\
16030 & 0.98 &  --- & --- &  0.09$\pm$0.69 &  ---  & --- & --- \\
16059 & 2.325\tablenotemark{c} &  --- & --- &  1.20$\pm$0.66 &  0.57$\pm$ 0.03  & --- & $>$\,0.77\tablenotemark{d} \\
16080 & 2.007\tablenotemark{c} &  5.2$\pm$1.7(6.0\arcsec) & --- &   0.69$\pm$0.54 &  0.34$\pm$ 0.03 & 0.63$\pm$0.07 &  -0.75 \\
16095 & 1.68 & 11.0$\pm$1.6(1.3\arcsec) & --- &  -0.65$\pm$0.50 &  ---  & --- & --- \\
16113 & 1.90 &  --- & --- &  0.72$\pm$0.52 &  ---  & --- & --- \\
16122 & 1.97 &  --- &  --- &   0.00$\pm$0.60 &  2.82$\pm$0.14  & 4.34$\pm$0.12 & -0.52 \\

22204 & 1.974\tablenotemark{c} & 12.9$\pm$1.8(2.7\arcsec) &   --- &   0.27$\pm$0.51 &  1.89$\pm$0.08  &  4.67$\pm$0.10 & -1.09 \\
22277 & 1.77 & 17.5$\pm$1.8(2.2\arcsec) &  --- &   0.86$\pm$0.6 &  1.57$\pm$ 0.07  & 4.79$\pm$0.11 & -1.34 \\
22303 & 2.34 &  6.6$\pm$1.5(1.4\arcsec) &   --- &  -1.05$\pm$0.64 &  0.26$\pm$ 0.02  & --- & $>$\,-0.17 \\

22467 & 0.70 &  9.1\,$\pm$\,1.5(3.3\arcsec) &  --- &  -0.73$\pm$0.70 &  ---  & --- & --- \\

22558 & 3.20 &  --- &  --- &  1.73$\pm$0.48 &  0.62$\pm$ 0.03 & 1.55$\pm$0.12 &  -1.11 \\

22661 & 1.80 &  --- & --- &  0.65$\pm$0.56 &  ---  & --- & --- \\
22699 & 2.59 &  --- & --- &   -0.31$\pm$0.63 &  ---  & 0.25$\pm$0.07 & $<$\,-1.23 \\
 \enddata
 
 \tablenotetext{a} {We use a format of Flux\,$\pm$\,1\,$\sigma$. The offset between the APEX source and 24\um\ source is given in brackets.. Large offsets are typically blended sources (see $^b$)}
 \tablenotetext{b}{Multiple 24\um\ sources. We give the deblended value (see \S\,\ref{sec_deblend} for details).}
\tablenotetext{c}{Redshift from optical or near-IR spectroscopy (see \S\,\ref{sec_opt1}). These are estimate to be accurate to $\pm$\,0.001. The rest of the redshifts are based on mid-IR features (see Paper\,I).}
\tablenotetext{d}{Potentially an inverted spectrum (see \S\,\ref{sec_lradio} for details). }

\end{deluxetable}

\begin{deluxetable}{cccccc}
\tablecolumns{6}
\tablewidth{5in}
\tabletypesize{\scriptsize}
\tablecaption{\label{table_det} 3$\sigma$ Detection statistics }
\tablehead{\colhead{} &  \colhead{All} & \multicolumn{2}{c}{strong-PAH}  &\multicolumn{2}{c}{weak-PAH}  \\
\cline{3-4}  \cline{5-6} \\
\colhead{} & \colhead{} &  \colhead{$z$\,$<$\,1.5} & \colhead{$z$\,$>$\,1.5} & \colhead{$z$\,$<$\,1.5} & \colhead{$z$\,$>$\,1.5}  }
\startdata
70\um\ & 60\% (29/48) & 100\% (6/6) & 50\% (4/8) & 57\% (4/7) & 56\% (15/27) \\
160\um\ & 23\% (11/48) & 50\% (3/6) & 50\% (4/8) & 14\% (1/7) & 11\% (3/27) \\
1.2\,mm & 16\% (7/44) & 0\% (0/2) & 50\% (4/8) & 0\% (0/7) & 11\% (3/27) \\
1.4\,GHz & 67\% (32/48) & 83\% (5/6) & 75\% (6/8) & 71\% (5/7) & 59\% (16/27) \\
610\,MHz & 48\% (23/48) & 50\% (3/6) & 50\% (4/8) & 43\% (3/7) & 48\% (13/27) \\
\enddata
\end{deluxetable}

\begin{deluxetable}{cccccccccc}
\tablecolumns{10}
\tablewidth{0pc}
\tabletypesize{\scriptsize}
\tablecaption{\label{table_fits} Fit results }
\tablehead{\colhead{MIPSID} &    \colhead{$\log(L_{\rm{1.4GHz}})$}  &\colhead{$\log(L_{\rm{AGN}})$\tablenotemark{a}} & \colhead{$\log(L_{\rm{SB}})$\tablenotemark{b}} & \colhead{$\log(L_{\rm{IR}})$\tablenotemark{c,d}} & \colhead{$T_{\rm{cold}}$}  &  \colhead{SFR} & \colhead{Radio} & \colhead{Optical}  \\
\colhead{} &   \colhead{W/Hz} & \colhead{$L_{\odot}$} & \colhead{$L_{\odot}$} &\colhead{$L_{\odot}$} &  \colhead{K}  & \colhead{$M_{\odot}/yr$} & \colhead{AGN} & \colhead{AGN} }
 \startdata
 strong-PAH & & & & & & & & \\
\hline \\
  283 & 23.84$\pm$0.16 & 11.3$\pm$0.1 & $<$11.9 & 11.3\,--\,12.0 & 40$\pm$5 & $<$ 110 &  no & no \\
  289 & $<$24.2 & 12.1$\pm$0.3 & 12.7$\pm$0.2 & 12.8$\pm$0.2 & 29$\pm$2 &  880$\pm$220 &  no & --- \\
  506 & 24.67$\pm$0.14 & 12.7$\pm$0.2 & $<$12.7 & 12.7\,--\,13.0 & 34$\pm$5 & $<$ 730 &  no & no \\
 8184 & 23.75$\pm$0.23 & 11.6$\pm$0.1& 12.3$\pm$0.2& 12.4$\pm$0.2 & 43$\pm$2 &  320$\pm$ 90 &  no & --- \\
 8207 & 23.89$\pm$0.08 & 11.4$\pm$0.1& 12.1$\pm$0.2& 12.2$\pm$0.2 & 40$\pm$1 &  210$\pm$ 60 &  no & no \\
 8493 & $<$24.2 & 11.8$\pm$0.3 & $<$12.4 & 11.8\,--\,12.5 & 32$\pm$4 & $<$ 450 &  no & --- \\
15928 & 25.27$\pm$0.01 & 12.6$\pm$0.1 & $<$12.1 & 12.6\,--\,12.7 & 25$\pm$4 & $<$ 200 & yes & yes \\
16144 & 24.46$\pm$0.43 & 12.3$\pm$0.2 & $<$12.8 & 12.3\,--\,12.9 & 30$\pm$3 & $<$ 920 &  no & --- \\
22404 & 23.43$\pm$0.16 & 11.1$\pm$0.1& 11.9$\pm$0.2& 12.0$\pm$0.2 & 35$\pm$2 &  130$\pm$ 40 &  no &  no \\
22482 & 24.81$\pm$0.06 & 12.4$\pm$0.2& 12.8$\pm$0.2& 12.9$\pm$0.2 & 60$\pm$3 &  900$\pm$360 & bor & --- \\
22530 & 24.83$\pm$0.22 & 12.4$\pm$0.3& 12.8$\pm$0.2& 12.9$\pm$0.2 & 33$\pm$2 &  990$\pm$400 & bor & no \\
22554 & 23.54$\pm$0.29 & 11.0$\pm$0.2 & $<$11.8 & 11.0\,--\,11.9 & 44$\pm$5 & $<$ 100 &  no &  no \\
22600 & $<$23.4 & 11.1$\pm$0.2 & $<$11.7 & 11.1\,--\,11.8 & 39$\pm$4 & $<$  90 &  no &  no \\
22651 & 24.92$\pm$0.03 & 12.2$\pm$0.1& 12.7$\pm$0.2& 12.8$\pm$0.2 & 30$\pm$2 &  860$\pm$210 & bor & --- \\
\hline \\
weak-PAH & & & & & & & & \\
\hline \\
   42 & $<$24.3 & 13.0$\pm$0.1 & $<$12.4 & 13.0\,--\,13.1 & 30$\pm$5 & $<$ 380 &  no & --- \\
   78 & $<$24.6 & 13.1$\pm$0.1 & $<$12.7 & 13.1\,--\,13.2 & 36$\pm$6 & $<$ 730 &  no & --- \\
  110 & 24.33$\pm$0.04 & 12.0$\pm$0.1 & $<$11.9 & 12.0\,--\,12.2 & 23$\pm$3 & $<$ 110 &  no & yes \\
  133 & 24.13$\pm$0.04 & 11.7$\pm$0.1 & $<$11.8 & 11.7\,--\,12.1 & 43$\pm$5 & $<$  90 &  no & yes \\
  180 & $<$24.5 & 12.9$\pm$0.2 & $<$12.6 & 12.9\,--\,13.1 & 35$\pm$5 & $<$ 700 &  no & --- \\
  227 & 24.61$\pm$0.06 & 12.6$\pm$0.1 & $<$12.2 & 12.6\,--\,12.8 & 28$\pm$4 & $<$ 290 & yes & yes \\
  279 & 24.24$\pm$0.08 & 11.6$\pm$0.1& 11.9$\pm$0.2& 12.1$\pm$0.2 & 40$\pm$3 &  140$\pm$ 60 &  no & no \\
  429 & $<$24.4 & 12.7$\pm$0.2 & $<$11.7 & 12.7\,--\,12.8 & 20$\pm$0 & $<$  90 &  no &  no \\
  464 & 24.42$\pm$0.40 & 12.5$\pm$0.2 & $<$12.1 & 12.5\,--\,12.7 & 26$\pm$7 & $<$ 220 &  no & --- \\
 8034 & 23.72$\pm$0.09 & 11.9$\pm$0.1 & $<$11.8 & 11.9\,--\,12.1 & 21$\pm$2 & $<$  90 &  no & --- \\
 8196 & $<$24.5 & 12.9$\pm$0.1 & $<$12.7 & 12.9\,--\,13.1 & 39$\pm$5 & $<$ 880 &  no & yes \\
 8242 & $<$24.5 & 12.7$\pm$0.2 & $<$12.9 & 12.7\,--\,13.1 & 34$\pm$3 & $<$1160 &  no & --- \\
 8245 & 24.76$\pm$0.05 & 12.8$\pm$0.1 & $<$12.7 & 12.8\,--\,13.1 & 39$\pm$6 & $<$ 790 & no & --- \\
 8268 & $<$23.4 & 11.4$\pm$0.1 & $<$11.6 & 11.4\,--\,11.8 & 20$\pm$0 & $<$  70 &  no & --- \\
 8327 & 25.88$\pm$0.02 & 12.7$\pm$0.1 & $<$12.7 & 12.7\,--\,13.0 & 36$\pm$5 & $<$ 750 & yes &  no \\
 8342 & 24.82$\pm$0.08 & 12.4$\pm$0.1& 12.1$\pm$0.2& 12.6$\pm$0.2 & 27$\pm$2 &  220$\pm$ 70 & bor & yes \\
15840 & $<$24.4 & 12.7$\pm$0.1 & $<$12.7 & 12.7\,--\,13.0 & 37$\pm$4 & $<$ 770 &  no & --- \\
15880 & 25.06$\pm$0.03 & 12.8$\pm$0.1& 11.9$\pm$0.1& 12.9$\pm$0.1 & 20$\pm$0 &  140$\pm$ 30 & yes & --- \\
15949 & 24.60$\pm$0.20 & 12.8$\pm$0.1 & $<$12.5 & 12.8\,--\,13.0 & 33$\pm$4 & $<$ 560 &  no & yes \\
15958 & 24.62$\pm$0.11 & 12.5$\pm$0.1 & $<$12.5 & 12.5\,--\,12.8 & 32$\pm$3 & $<$ 520 &  no & --- \\
15977 & 24.81$\pm$0.08 & 12.6$\pm$0.1& 12.8$\pm$0.2& 13.0$\pm$0.2 & 56$\pm$2 &  920$\pm$230 & bor & --- \\
16030 & $<$23.6 & 11.5$\pm$0.1 & $<$11.8 & 11.5\,--\,12.0 & 22$\pm$2 & $<$ 110 &  no &  no \\
16059 & 25.23$\pm$0.04 & 12.7$\pm$0.1 & $<$12.6 & 12.7\,--\,12.9 & 33$\pm$5 & $<$ 610 & yes & yes \\
16080 & 24.88$\pm$0.05 & 12.6$\pm$0.1 & $<$12.3 & 12.6\,--\,12.8 & 30$\pm$6 & $<$ 320 & bor & no \\
16095 & $<$24.1 & 12.6$\pm$0.1 & $<$12.2 & 12.6\,--\,12.7 & 27$\pm$5 & $<$ 230 &  no & --- \\
16113 & $<$24.2 & 12.2$\pm$0.2 & $<$12.5 & 12.2\,--\,12.7 & 32$\pm$4 & $<$ 470 &  no & --- \\
16122 & 25.68$\pm$0.01 & 12.4$\pm$0.2 & $<$12.5 & 12.4\,--\,12.8 & 32$\pm$4 & $<$ 480 & yes & --- \\
22204 & 25.78$\pm$0.01 & 12.9$\pm$0.1 & $<$12.4 & 12.9\,--\,13.0 & 31$\pm$5 & $<$ 380 & yes & yes \\
22277 & 25.69$\pm$0.01 & 12.8$\pm$0.2 & $<$11.8 & 12.8\,--\,12.9 & 20$\pm$0 & $<$  90 & yes & --- \\
22303 & 24.90$\pm$0.09 & 12.9$\pm$0.1 & $<$12.5 & 12.9\,--\,13.0 & 32$\pm$6 & $<$ 500 & yes & --- \\
22467 & $<$23.2 & 11.1$\pm$0.1 & $<$11.5 & 11.1\,--\,11.7 & 39$\pm$4 & $<$  50 &  no & --- \\
22558 & 25.83$\pm$0.02 & 13.3$\pm$0.2 & $<$12.0 & 13.3\,--\,13.3 & 20$\pm$0 & $<$ 160 & yes & --- \\
22661 & $<$24.2 & 12.2$\pm$0.1 & $<$12.4 & 12.2\,--\,12.6 & 31$\pm$4 & $<$ 420 &  no & --- \\
22699 & $<$24.5 & 12.6$\pm$0.1 & $<$12.8 & 12.6\,--\,13.0 & 39$\pm$4 & $<$1060 &  no & --- \\
\enddata
 
 \tablenotetext{a}{The integrated 3\,---\,1000\um\ luminosity of the AGN component alone. }
 \tablenotetext{b}{The AGN-subtracted $L_{\rm{IR}}$. Includes both cold dust and PAH emission. It is an upper limit for all sources without 160\um\ detections. In these cases, the 2\,$\sigma$ limits are used in the fits.}
 \tablenotetext{c}{The total integrated 3\,--\,1000\um\ luminosity.} 
 \tablenotetext{d}{For 160\um-undetected sources,  we take $L_{\rm{IR}}$ to be in the range from $L_{\rm{AGN}}$ (i.e. mid-IR continuum) alone to $L_{\rm{AGN}}$ and the upper limit on $L_{\rm{SB}}$.}
 
\end{deluxetable}


\begin{deluxetable}{ccccccccc}
\tablecolumns{9}
\tablewidth{0in}
\tabletypesize{\scriptsize}
\tablecaption{\label{table_keck} Near-IR spectroscopy results }
\tablehead{\colhead{MIPSID} &  \colhead{$z_{\rm{opt}}$}  &\colhead{FWHM(H$\alpha$)}  & \colhead{$\log(L_{\rm{H}\alpha})$} & \colhead{FWHM(O{\sc iii})} & \colhead{$\log(L_{\rm{OIII}})$} & \colhead{log(N{\sc ii}/H$\alpha$)} & \colhead{log(O{\sc iii}/H$\beta$)} & Classification \\
\colhead{} & \colhead{} &  \colhead{km/s} & \colhead{\lsun} &   \colhead{km/s} & \colhead{\lsun} & \colhead{} & \colhead{} & \colhead{} }
\startdata
  429 &  2.213 &   918 &  8.61 &    --- & --- & $<$\,-1.00 & --- &  Starburst \\
   506 &  2.470 &   743 &  8.38 &    --- & --- & -0.46 & --- &   Starburst \\
 8196 &  2.586 &   562 &  8.84 &   702 &  9.02 & -0.91 &  0.69 &  Composite \\
 8327 &  2.441 &   971 &  8.92 &   634 &  8.94 & --- &  0.54 &  Composite \\ 
  8342 &  1.562 &   771 &  7.61 &    --- & --- & -0.18 & --- &  Composite \\
15949 &  2.122 &  2189 &  9.21 &  1304 &  8.93 & -1.44 &  $>$\,1.06 &   AGN \\
16059 &  2.326 &   666 &  8.21 &   671 &  8.56 & -0.64 & $>$\,1.70 &  AGN \\
16080 &  2.007 &  1359 &  9.13 &  1155 &  8.88 & -0.88 & --- &  Starburst\tablenotemark{a} \\
16144 &  2.131 &   522 &  8.68 &    --- & --- & -0.40 & --- &  Starburst \\
22204 &  1.974 &  -- &  -- &  -- &  -- & --- & --- &  Blended?\tablenotemark{b} \\
22530 &  1.952 &   461 &  8.37 &   --- &  --- & -0.86 & --- &  Starburst \\

\enddata

\tablenotetext{a}{The line ratio is starburst-like. However, this source shows the asymmetric [OIII] profile typical of AGN (see \S\,\ref{sec_asymm}). }
\tablenotetext{b}{This source is classified as a Type 2 AGN based on its UV spectrum \citep{ams07}.}
\end{deluxetable}


\begin{thebibliography}{}

\bibitem[\protect\citeauthoryear{Alexander et al.}{2005}]{alexander05} Alexander, D.~M., 
Bauer, F.~E., Chapman, S.~C., Smail, I., Blain, A.~W., Brandt, W.~N., \& 
Ivison, R.~J.\ 2005, \apj, 632, 736 
\bibitem[\protect\citeauthoryear{Appleton et al.}{2004}]{appleton04} Appleton, P.~N., et al.\ 2004, \apjs, 154, 147
\bibitem[\protect\citeauthoryear{Armus et al.}{2007}]{armus07} Armus, L. et al. 2007, \apj, 656, 148

\bibitem[\protect\citeauthoryear{Bavouzet et al.}{2008}]{nb07} Bavouzet, N., Dole, H., LeFloch, E., Caputi, K.I., Lagache, G., Kochanek, C.\ 2008, A\&A, 479, 83
\bibitem[\protect\citeauthoryear{Beelen et al.}{2006}]{beelen06} Beelen, A., Cox, P., Benford, D.~J., Dowell, C.~D., Kov{\'a}cs, A., Bertoldi, F., Omont, A., \& Carilli, C.~L.\ 2006, \apj, 642, 694 

\bibitem[\protect\citeauthoryear{Bondi et al.}{2007}]{bondi07} Bondi, M., et al.\ 2007, \aap, 463, 519
\bibitem[\protect\citeauthoryear{Brand et al.}{2007}]{brand07} Brand, K., et al.\ 2007, \apj, 663, 204 
\bibitem[\protect\citeauthoryear{Brandl et al.}{2006}]{brandl06} Brandl, B.~R., et al.\ 2006, \apj, 653, 1129 

\bibitem[\protect\citeauthoryear{Caputi et al.}{2007}]{caputi07} Caputi, K., Lagache, G., Yan, L., et al. 2007, ApJ, 660, 97
\bibitem[\protect\citeauthoryear{Chapman et al.}{2004}]{chapman04} Chapman, S.C., Smail, I., Windhorst, R., Muxlow, T., Ivison, R.J., 2004, \apj, 611, 732  
\bibitem[\protect\citeauthoryear{Chary \& Elbaz}{2001}]{ce01} Chary, R.R., Elbaz, D., 2001, ApJ, 556, 562
\bibitem[\protect\citeauthoryear{Chiar \& Tielens}{2006}]{chiar06} Chiar, J.~E., \& Tielens, A.~G.~G.~M.~2006, \apj, 637, 774
\bibitem[\protect\citeauthoryear{Condon}{1992}]{condon92} Condon, J.~J.\ 1992, \araa, 30, 575 
\bibitem[\protect\citeauthoryear{Condon et al.}{2002}]{condon02} Condon, J.~J., Helou, G., \& Jarrett, T.~H.\ 2002, \aj, 123, 1881 
\bibitem[\protect\citeauthoryear{Condon et al.}{2003}]{condon03} Condon, J.J, Cotton, W.D., Yin Q.F., Shupe, D.L., Storrie-Lombardi, L.J., Helou, G., Soifer, T., Werner, M.W. ~2003, AJ, 125, 2411

\bibitem[\protect\citeauthoryear{Coppin et al.}{2006}]{coppin06} Coppin, K., et al.\ 
2006, \mnras, 372, 1621 

\bibitem[\protect\citeauthoryear{Dasyra et al.}{2008}] {dasyra08} Dasyra, K., et al. 2008, ApJ, in press, (arXiv:0802.1050)
\bibitem[\protect\citeauthoryear{Desai et al.}{2007}]{desai07} Desai, V., Armus, L., et al., 2007,\apj, 669, 810
 \bibitem[\protect\citeauthoryear{Dole et al.}{2004a}]{dole04_counts} Dole, H., et al.\ 2004a, 
\apjs, 154, 87 
\bibitem[\protect\citeauthoryear{Dole et al.}{2004}]{dole04_confusion} Dole, H., et al.\ 2004b, \apjs, 154, 93 
\bibitem[\protect\citeauthoryear{Downes et al.}{1986}]{downes86} Downes, A.~J.~B., 
Peacock, J.~A., Savage, A., \& Carrie, D.~R.\ 1986, \mnras, 218, 31 
\bibitem[\protect\citeauthoryear{Dunne \& Eales}{2001}]{dunne01} Dunne, L., \& Eales, S.~A.\ 2001, \mnras, 327, 697 

\bibitem[\protect\citeauthoryear{Efstathiou \& Rowan-Robinson}{1995}]{e_mrr} Efstathiou, A., \& Rowan-Robinson, M.\ 1995, \mnras, 273, 649 
\bibitem[\protect\citeauthoryear{Elvis et al.}{1994}]{elvis94} Elvis, M., et al.\ 1994, \apjs, 95, 1

\bibitem[\protect\citeauthoryear{Faber et al.}{2003}]{faber} Faber, S.~M., et al.\ 
2003, \procspie, 4841, 1657 
\bibitem[\protect\citeauthoryear{Farrah et al.}{2002}]{farrah02} Farrah, D., Serjeant, S., Efstathiou, A., Rowan-Robinson, M., Verma, A.~2002, \mnras, 335,1163
\bibitem[\protect\citeauthoryear{Farrah et al.}{2003}]{farrah03} Farrah, D., Afonso, J., 
Efstathiou, A., Rowan-Robinson, M., Fox, M., \& Clements, D.\ 2003, \mnras, 
343, 585 
\bibitem[\protect\citeauthoryear{Feigelson \& Nelson}{1985}]{fn85} Feigelson, E.~D., 
\& Nelson, P.~I.\ 1985, \apj, 293, 192 
\bibitem[\protect\citeauthoryear{Frayer et al.}{2006a}]{frayer06} Frayer, D.~T., et al.\ 2006a, \aj, 131, 250
\bibitem[\protect\citeauthoryear{Frayer et al.}{2006b}]{frayer06_counts70} Frayer, D.~T., et al.\ 
2006b, \apjl, 647, L9 
\bibitem[\protect\citeauthoryear{Fritz et al.}{2006}]{fritz06} Fritz, J., Franceschini, A., \& Hatziminaoglou, E.\ 2006, \mnras, 366, 767 

\bibitem[\protect\citeauthoryear{Garn et al.}{2007}]{garn07} Garn, T., Green, D.A., Hales, S.E.G., Riley, J.M., Alexander,P.~2007, MNRAS, 376, 1251
\bibitem[\protect\citeauthoryear{Genzel et al.}{1998}]{genzel98} Genzel, R., et al.\ 
1998, \apj, 498, 579 

 
\bibitem[\protect\citeauthoryear{Heckman et al.}{1981}]{heckman81} Heckman, T.~M., Miley, 
G.~K., van Breugel, W.~J.~M., \& Butcher, H.~R.\ 1981, \apj, 247, 403 
\bibitem[\protect\citeauthoryear{Helou et al.}{1985}]{helou85} Helou, G., Soifer, B.~T., 
\& Rowan-Robinson, M.\ 1985, ApJL, 298, 7 
\bibitem[\protect\citeauthoryear{Helou et al.}{1988}]{helou88} Helou, G., Khan, I.R., Malek, L., Boehmer, L.~1988, ApJS, 68, 151
\bibitem[\protect\citeauthoryear{Hildebrand}{1983}]{hildebrand83} Hildebrand, R. H. 1983, QJRAS, 24, 267
\bibitem[\protect\citeauthoryear{Hodapp et al.}{2003}]{hodapp} Hodapp, K.~W., et al.\ 
2003, \pasp, 115, 1388 

\bibitem[\protect\citeauthoryear{Holland et al.}{2006}]{scuba2} Holland, W., et al.\ 
2006, \procspie, 6275,  45

\bibitem[\protect\citeauthoryear{Hopkins et al.}{2007}]{hopkins07} Hopkins, P.~F., 
Richards, G.~T., \& Hernquist, L.\ 2007, \apj, 654, 731 
\bibitem[\protect\citeauthoryear{Houck et al.}{2005}]{houck05} Houck, J.~R., et al.~2005, \apjl, 622, 105
\bibitem[\protect\citeauthoryear{Houck et al.}{2004}]{houck04} Houck, J.~R., et al.\ 
2004, \procspie, 5487, 62 
\bibitem[\protect\citeauthoryear{Huynh et al.}{2007a}]{huynh07a} Huynh, M.~T., Jackson, C.~A., \& Norris, R.~P.\ 2007a, \aj, 133, 1331
\bibitem[\protect\citeauthoryear{Huynh et al.}{2007b}]{huynh07b} Huynh, M.~T., Pope, A., Frayer, D.~T., \& Scott, D.\ 2007, \apj, 659, 305 
\bibitem[\protect\citeauthoryear{Kauffmann et al.}{2003}]{kauffmann03} Kauffmann, G., et 
al.\ 2003, \mnras, 346, 1055 
\bibitem[\protect\citeauthoryear{Kelson}{2003}]{kelson03} Kelson, D.~D.\ 2003, \pasp, 115, 688 
\bibitem[\protect\citeauthoryear{Kennicutt}{1998}]{ken98} Kennicutt, R.~C., Jr.\ 1998, \araa, 36, 189
\bibitem[\protect\citeauthoryear{Kewley et al.}{2006}]{kewley06} Kewley, L.~J., Groves, 
B., Kauffmann, G., \& Heckman, T.\ 2006, \mnras, 372, 961 
\bibitem[\protect\citeauthoryear{Klaas et al.}{2001}]{klaas01} Klaas, U., et al.\ 2001, \aap, 379, 823 
\bibitem[\protect\citeauthoryear{Kov{\'a}cs et al.}{2006}]{kovacs06} Kov{\'a}cs, A., 
Chapman, S.~C., Dowell, C.~D., Blain, A.~W., Ivison, R.~J., Smail, I., \& 
Phillips, T.~G.\ 2006, \apj, 650, 592 
\bibitem[\protect\citeauthoryear{Kreysa et al.}{1998}]{kreysa98} Kreysa, E., et al.\ 
1998, \procspie, 3357, 319 
\bibitem[\protect\citeauthoryear{Krisciunas et al.}{1987}]{maunakea_ext} Krisciunas, K., et 
al.\ 1987, \pasp, 99, 887 
 
\bibitem[\protect\citeauthoryear{Lacy et al.}{2007}]{lacy07_letter} Lacy, M., Sajina, A., Petric, A.O., Seymour, N., Canalizo, G., Ridgway, S.E., Armus L., Storrie-Lombardi, L.~2007, ApJL, 669, 61 
\bibitem[\protect\citeauthoryear{Le Floc'h et al.}{2005}]{lefloch05} Le Floc'h, E., et al.\ 2005, \apj, 632, 169 
\bibitem[\protect\citeauthoryear{Levenson et al.}{2007}]{levenson07} Levenson, N.A., Sirocky, M.M., Hao, L., Spoon, H.W.W., Marshall, J.A., Elitzur, M., Houck, J.R., 2007, ApJL, 654,45

\bibitem[\protect\citeauthoryear{Lonsdale et al.}{2003}]{lonsdale03} Lonsdale, C.~J., et 
al.\ 2003, \pasp, 115, 897 
\bibitem[\protect\citeauthoryear{Lonsdale et al.}{2006}]{lonsdale06} Lonsdale, C.~J., 
Farrah, D., \& Smith, H.~E.\ 2006, Astrophysics Update 2, 285 

\bibitem[\protect\citeauthoryear{Lutz et al.}{1998}]{lutz98} Lutz, D., Spoon, H.~W.~W., 
Rigopoulou, D., Moorwood, A.~F.~M., \& Genzel, R.\ 1998, ApJL, 505, 103 
\bibitem[\protect\citeauthoryear{Lutz et al.}{2005}]{lutz05} Lutz, D., Yan, L., Armus,  L., Helou, G., Tacconi, L.~J., Genzel, R., \& Baker, A.~J.\ 2005, ApJL,  632, 13 
\bibitem[\protect\citeauthoryear{Magliocchetti et al.}{2008}]{manuela08} Magliocchetti, 
M., Andreani, P., \& Zwaan, M.~A.\ 2008, ApJ, 383, 479

\bibitem[\protect\citeauthoryear{Makovoz \& Marleau}{2005}]{makovoz05} Makovoz, D., \& Marleau, F.~R.\ 2005, \pasp, 117, 1113
\bibitem[\protect\citeauthoryear{Marshall et al.}{2007}]{marshall07} Marshall, J.~A., 
Herter, T.~L., Armus, L., Charmandaris, V., Spoon, H.~W.~W., Bernard-Salas, 
J., \& Houck, J.~R.\ 2007, \apj, 670, 129

\bibitem[\protect\citeauthoryear{Mart{\'{\i}}nez-Sansigre et al.}{2006}]{alejo06b} Mart{\'{\i}}nez-Sansigre, A., Rawlings, S., Garn, T., Green, D.~A., Alexander, P., Kl{\"o}ckner, H.-R., \& Riley, J.~M.\ 2006, \mnras, 373, L80 
\bibitem[\protect\citeauthoryear{Mart{\'{\i}}nez-Sansigre et al.}{2008}]{ams07} Mart{\'{\i}}nez-Sansigre, A., Lacy, M., Sajina, A., Rawlings, S.~2008, \apj, 674, 676
\bibitem[\protect\citeauthoryear{McLean et al.}{1995}]{mclean} McLean, I.~S., Becklin, 
E.~E., Figer, D.~F., Larson, S., Liu, T., \& Graham, J.\ 1995, \procspie, 2475, 350 
\bibitem[\protect\citeauthoryear{Miller \& Owen}{2001}]{mo01} Miller, N.~A., \& Owen, 
F.~N.\ 2001, \aj, 121, 1903 
\bibitem[\protect\citeauthoryear{Morganti et al.}{2004}]{morganti04} Morganti, R., Garrett, M.~A., Chapman, S., Baan, W., Helou, G., \& Soifer, T.\ 2004, \aap, 424, 371 
\bibitem[\protect\citeauthoryear{Murphy et al.}{2006}]{murphy06} Murphy, E.~J., et al.\ 
2006, ApJL, 651, 111 
\bibitem[\protect\citeauthoryear{Netzer et al.}{2006}]{netzer06} Netzer, H., Mainieri, 
V., Rosati, P., \& Trakhtenbrot, B.\ 2006, \aap, 453, 525 

\bibitem[\protect\citeauthoryear{Oke et al.}{1995}]{oke} Oke, J.~B., et al.\ 1995, 
\pasp, 107, 375 

\bibitem[\protect\citeauthoryear{Osterbrock}{1989}]{osterbrock} Osterbrock, D. E. 1989, Astrophysics of gaseous nebulae and active galactic 
nuclei (Research supported by the University of California, John Simon Guggenheim Memorial Foundation, University of Minnesota, et al. Mill 
Valley, CA, University Science Books, 1989, 422 p.)

\bibitem[\protect\citeauthoryear{Polletta et al.}{2008}]{polletta07} Polletta, M., Weedman, D., H\"onig, S., Lonsdale, C.J., Smith, H.E., Houck, J. 2008, \apj, 675, 960

\bibitem[\protect\citeauthoryear{Pope et al.}{2006}]{pope06} Pope, A., et al.~2006, MNRAS, 370, 1185
\bibitem[\protect\citeauthoryear{Pope et al.}{2008}]{pope07} Pope, A., et al.~2008,  ApJ, 675, 1171

\bibitem[\protect\citeauthoryear{Rice et al.}{2006}]{rice06} Rice, M.~S., Martini, P., Greene, J.~E., Pogge, R.~W., Shields, J.~C., Mulchaey, J.~S., \& Regan, M.~W.\ 2006, \apj, 636, 654 
\bibitem[\protect\citeauthoryear{Richards et al.}{2006}]{richards06} Richards, G.~T., et al.\ 2006, \aj, 131, 2766 
\bibitem[\protect\citeauthoryear{Rieke et al.}{2004}]{rieke04} Rieke, G.~H., et al.\ 2004, \apjs, 154, 25
\bibitem[\protect\citeauthoryear{Risaliti et al.}{2006}]{risaliti06} Risaliti, et al. 2006, MNRAS, 365, 303
\bibitem[\protect\citeauthoryear{Roussel et al.}{2006}]{roussel06} Roussel, H, Helou, G., Smith, J.D., et al. 2006, ApJ, 646, 841
\bibitem[\protect\citeauthoryear{Sajina et al.}{2006}]{me06} Sajina, A., Scott, D., Dennefeld, M., Dole, H., Lacy, M., Lagache, G.,~2006, MNRAS, 369, 936
\bibitem[\protect\citeauthoryear{Sajina et al.}{2007a}]{me07} Sajina, A., Yan, L., Armus, L., Choi, P., Fadda D., Helou, G., Spoon, H.~2007a, ApJ, 664, 713
\bibitem[\protect\citeauthoryear{Sajina et al.}{2007b}]{rl_letter} Sajina, A., Yan, L., Lacy, M., Huynh, M., 2007b, ApJL, 667, 17
\bibitem[\protect\citeauthoryear{Sanders et al.}{1988}]{sanders88} Sanders, D.~B., Soifer, 
B.~T., Elias, J.~H., Neugebauer, G., \& Matthews, K.\ 1988, \apjl, 328, L35
\bibitem[\protect\citeauthoryear{Sanders \& Mirabel}{1996}]{sm96} Sanders D.B., Mirabel I.F.,~1996, ARA$\&$A, 34, 749

\bibitem[\protect\citeauthoryear{Smith et al.}{2007}]{jdsmith06} Smith, J.D., Draine, B.T., Dale, D., et al. 2007, ApJ, 656,770
\bibitem[\protect\citeauthoryear{Soifer et al.}{1987}]{soifer87} Soifer, B.~T., Neugebauer, G., \& Houck, J.~R.\ 1987, \araa, 25, 187
\bibitem[\protect\citeauthoryear{Spergel et al.}{2003}]{spergel03} Spergel, D.~N., et al.\ 
2003, \apjs, 148, 175 
\bibitem[\protect\citeauthoryear{Stansberry et al.}{2007}]{mips160_calib} Stansberry, J.A., et al. 2007, PASP, 119, 1038
\bibitem[\protect\citeauthoryear{Stern et al.}{2000}]{stern00} Stern, D., Djorgovski, S.~G., Perley, R.~A., de Carvalho, R.~R., \& Wall, J.~V.\ 2000, \aj, 119, 
1526
\bibitem[\protect\citeauthoryear{Sulentic \& Marziani}{1999}]{sm99} Sulentic, J.~W., \& Marziani, P.\ 1999, \apjl, 518, L9 
\bibitem[\protect\citeauthoryear{Tacconi et al.}{2006}]{tacconi06} Tacconi, L.~J., et al.\ 
2006, \apj, 640, 228
\bibitem[\protect\citeauthoryear{Takeuchi et al.}{2005}]{tak05} Takeuchi, L.~J., et al.\ 
2005, A\&A, 432, 423
\bibitem[\protect\citeauthoryear{Tran et al.}{2001}]{tran01} Tran, Q.~D., et al.\ 2001, \apj, 552, 527 
\bibitem[\protect\citeauthoryear{Valiante et al.}{2007}]{valiante07} Valiante, E., Lutz, 
D., Sturm, E., Genzel, R., Tacconi, L.~J., Lehnert, M.~D., \& Baker, A.~J.\ 
2007, \apj, 660, 1060 

\bibitem[\protect\citeauthoryear{Weedman et al.}{2006}]{weedman06} Weedman, D., et al.\ 
2006, \apj, 653, 101 
\bibitem[\protect\citeauthoryear{Werner et al.}{2004}]{werner04} Werner, M.~W., et al.\ 2004, \apjs, 154, 1
\bibitem[\protect\citeauthoryear{Yan et al.}{2004a}]{yan04aj} Yan, L., Thompson, D., \& 
Soifer, B.~T.\ 2004a, \aj, 127, 1274 
\bibitem[\protect\citeauthoryear{Yan et al.}{2004b}]{yan04} Yan, L., et al.~2004b, ApJ, 154, 60

\bibitem[\protect\citeauthoryear{Yan et al.}{2005}]{yan05} Yan, L., et al.~2005, \apj, 628, 604 
\bibitem[\protect\citeauthoryear{Yan et al.}{2007}]{yan07} Yan, L., et al.~2007, \apj, 658, 778
\bibitem[\protect\citeauthoryear{Yun et al.}{2001}]{yun01} Yun, M.S., Reddy, N.A., Condon, J.J.~2001, \apj, 554, 803
\bibitem[\protect\citeauthoryear{Zakamska et al.}{2003}]{zakamska03} Zakamska, N.~L., et 
al.\ 2003, \aj, 126, 2125 
\bibitem[\protect\citeauthoryear{Zhou et al.}{2006}]{zhou06} Zhou, H., Wang, T., Yuan, 
W., Lu, H., Dong, X., Wang, J., \& Lu, Y.\ 2006, \apjs, 166, 128 

\end{thebibliography}
\end{document}